\documentclass[manuscript,screen]{acmart}
\AtBeginDocument{%
  }

\usepackage{enumitem}
\usepackage[table]{xcolor}  
\usepackage{colortbl}
\usepackage{pifont}
\usepackage[most]{tcolorbox}
\usepackage{natbib}
\usepackage[normalem]{ulem}
\usepackage{multirow}
\usepackage{subcaption}
\usepackage[framemethod=tikz]{mdframed}
\newcommand{\uparrowsymbol}{$\uparrow$}
\newcommand{\downarrowsymbol}{$\downarrow$}

\definecolor{c4}{cmyk}{0.6765,0.2017,0,0.0667}
\newtcbox{\hlthirdtab}{on line, rounded corners, box align=base, colback=c4!10,colframe=white,size=fbox,arc=3pt, before upper=\strut, top=-2pt, bottom=-4pt, left=-2pt, right=-2pt, boxrule=0pt}

\mdfdefinestyle{customstyle}{
  linecolor=gray!100,
  linewidth=3pt,
  innerleftmargin=3pt,
  topline=false,
  rightline=false,
  bottomline=false,
  leftline=true,
  innerrightmargin=3pt,
  innertopmargin=3pt,
  innerbottommargin=3pt,
  backgroundcolor=gray!15
}

\newenvironment{custommdframed}
  {\begin{mdframed}[style=customstyle, skipabove=3mm]}
  {\end{mdframed}}

\newcommand{\cmark}{\textcolor{green!60!black}{\ding{51}}} 
\newcommand{\xmark}{\textcolor{red}{\ding{55}}} 
\newcommand{\percent}[1]{{\small\hlthirdtab{#1}}}

\newcommand{\ourdata}{\textsc{CodeProjectEval}}
\newcommand{\ourmethod}{\textsc{ProjectGen}}

\setcopyright{acmlicensed}
\copyrightyear{2018}
\acmYear{2018}
\acmDOI{XXXXXXX.XXXXXXX}
\acmConference[Conference acronym 'XX]{Make sure to enter the correct
  conference title from your rights confirmation email}{June 03--05,
  2018}{Woodstock, NY}
\acmISBN{978-1-4503-XXXX-X/2018/06}

\begin{document}

\title{Towards Realistic Project-Level Code Generation via Multi-Agent Collaboration and Semantic Architecture Modeling}

\author{Qianhui Zhao}
\email{zhaoqianhui@buaa.edu.cn}
\affiliation{%
  \institution{Beihang University}
  \city{Beijing}
  \country{China}
}

\author{Li Zhang}
\affiliation{%
  \institution{Beihang University}
  \city{Beijing}
  \country{China}
}
\email{lily@buaa.edu.cn}

\author{Fang Liu}
\authornote{Corresponding author.}
\affiliation{%
  \institution{Beihang University}
  \city{Beijing}
  \country{China}
}
\email{fangliu@buaa.edu.cn}

\author{Junhang Cheng}
\affiliation{%
  \institution{Beihang University}
  \city{Beijing}
  \country{China}
}
\email{chengjunhang7@gmail.com}

\author{Chengru Wu}
\affiliation{%
  \institution{Beihang University}
  \city{Beijing}
  \country{China}
}
\email{chengru_wu@buaa.edu.cn}

\author{Junchen Ai}
\affiliation{%
  \institution{Beihang University}
  \city{Beijing}
  \country{China}
}
\email{junchen_ai@buaa.edu.cn}

\author{Qiaoyuanhe Meng}
\affiliation{%
  \institution{Beihang University}
  \city{Beijing}
  \country{China}
}
\email{mengqiaoyuanhe@buaa.edu.cn}

\author{Lichen Zhang}
\affiliation{%
  \institution{Independent}
  \city{Beijing}
  \country{China}
}
\email{lczhang9653@gmail.com}

\author{Xiaoli Lian}
\affiliation{%
  \institution{Beihang University}
  \city{Beijing}
  \country{China}
}
\email{lianxiaoli@buaa.edu.cn}

\author{Shubin Song}
\affiliation{%
  \institution{Software IDE Innovation Lab, Huawei Central Software Institute}
  \country{China}
}
\email{songshubin2@huawei.com}

\author{Yuanping Guo}
\affiliation{%
  \institution{Software IDE Innovation Lab, Huawei Central Software Institute}
  \country{China}
}
\email{guoyuanping1@huawei.com}

\renewcommand{\shortauthors}{Zhao et al.}

\begin{abstract}
In recent years, Large Language Models (LLMs) have achieved remarkable progress in automated code generation. In real-world software engineering, the growing demand for rapid iteration and continuous delivery underscores the importance of project-level code generation, where LLMs are expected to generate complete software projects directly from complex user requirements.
This task, however, presents greater challenges due to larger input scales, multi-modal information, and the need for holistic architectural reasoning. Although existing studies have made initial explorations, they still face key limitations, including unrealistic datasets and unreliable evaluation metrics that fail to reflect real-world complexity, the semantic gap between human-written requirements and machine-interpretable structures, and difficulties in managing hierarchical dependencies and maintaining quality throughout the generation process.
To address these limitations, we first introduce \ourdata{}, a project-level code generation dataset built from 18 real-world repositories with 12.7 files and 2,388.6 lines of code per task on average, supplemented with documentation and executable test cases for automatic evaluation. We further propose \ourmethod{}, a multi-agent framework that decomposes projects into architecture design, skeleton generation, and code filling stages with iterative refinement and memory-based context management. Within this framework, we introduce the Semantic Software Architecture Tree (SSAT), a structured and semantically rich representation that effectively bridges user requirements and source code implementation. Experiments show that \ourmethod{} achieves state-of-the-art performance, passing 52/124 test cases on the small-scale project-level code generation dataset DevBench, a 57\% improvement over the baseline approaches,
and 310 test cases on \ourdata{}, representing an improvement of roughly tenfold compared to the baselines.
\end{abstract}



\received{20 February 2007}
\received[revised]{12 March 2009}
\received[accepted]{5 June 2009}

\maketitle

\section{Introduction}

Driven by agile iteration and rapid delivery, modern software engineering faces shorter cycles and higher complexity, revealing the inefficiency of manual coding and maintenance.
Code generation constitutes a fundamental capability in contemporary software engineering, as it enables the automation of repetitive and labor-intensive programming tasks, thereby enhancing productivity. 
Recent advances in Large Language Models (LLMs), such as Claude \cite{anthropic2025claude4}, DeepSeek \cite{liu2024deepseekv3, guo2025deepseekr1}, Gemini \cite{team2023gemini, team2024gemini}, and GPT \cite{hurst2024gpt, achiam2023gpt}, have profoundly reshaped the landscape of intelligent software development, demonstrating remarkable proficiency in understanding natural language and producing high-quality code.
In parallel, a new class of vibe coding tool, such as Cursor\footnote{\url{https://www.cursor.com}}, Windsurf\footnote{\url{https://windsurf.com}}, and Claude Code\footnote{\url{https://www.claude.com/product/claude-code}}, has emerged, integrating LLM-based reasoning and in-context interaction into IDEs to support interactive and repository-aware code generation. 
These tools have made significant progress toward bridging human–AI collaboration by enabling iterative refinement, context retention, and partial project synthesis. 
However, most existing studies and tools focus on function-level or file-level code generation \cite{austin2021program, chen2021evaluating, wang2024rlcoder, zhang2024codeagent}, while research on \textit{project-level code generation} remains scare, which refers to the task of generating complete projects directly from complicated user requirements.
Despite the practical utility, their performance remains limited when confronted with complex, large-scale code generation tasks that require coherent architecture design, cross-module dependency reasoning, and long-range context understanding.

\begin{figure}[t]
  \centering
  \setlength{\abovecaptionskip}{0.1cm}
  \includegraphics[width=\linewidth]{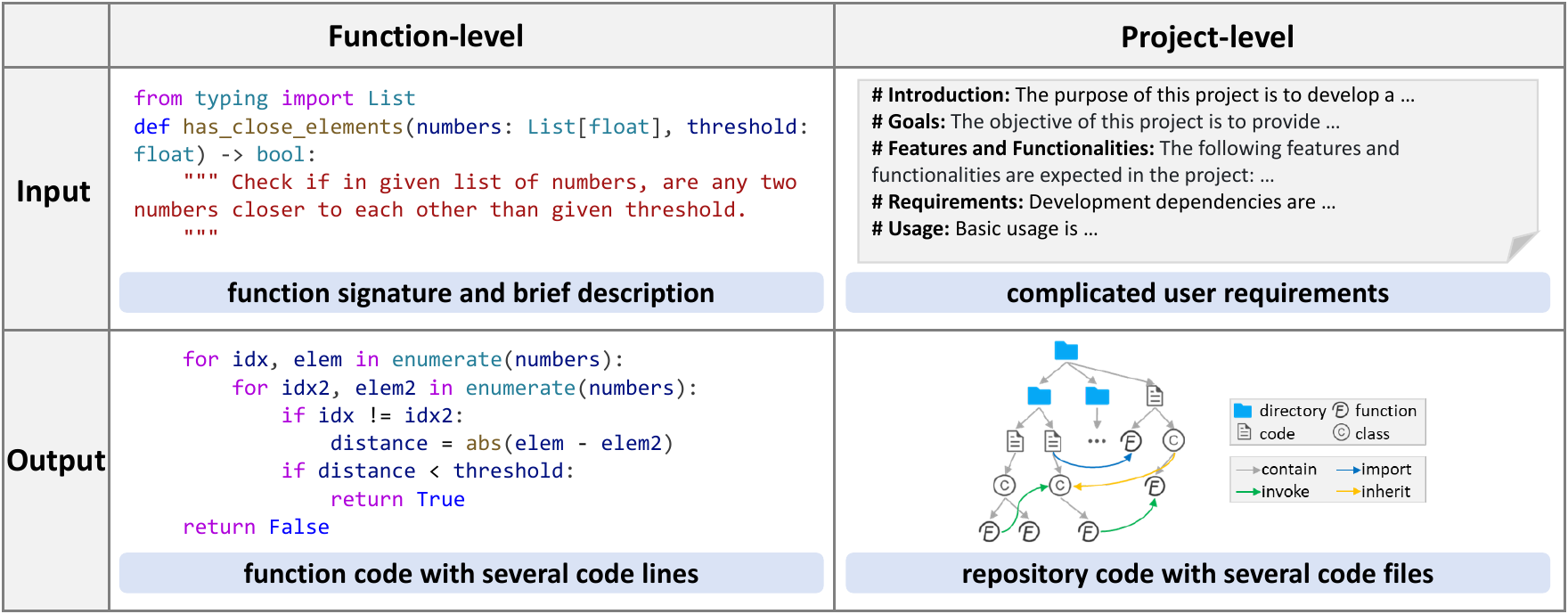}
  \caption{Task input and output comparison between function-level and project-level code generation.}
  \label{fig:task_description}
  \vspace{-0.3cm}
\end{figure}

Project-level code generation differs fundamentally in both scope and complexity from the more established function-level and file-level code generation.  
As shown in Figure \ref{fig:task_description}, while the latter focuses on producing isolated code snippets or single files based on localized contexts, project-level generation operates on a significantly larger scale. 
The input of project-level code generation tends to be much longer with diverse sources of information. It typically includes a project title, functional description, features, dependencies, usage examples, FAQs, \textit{etc} \cite{zan2024codes}.
The expected size of the produced code is also way larger than its descriptions.
For the generation complexity, project-level code generation requires LLMs to engage in holistic reasoning about the global architectural structure, which involves intricate planning for architecture design, module organization, and cross-file dependencies. Consequently, project-level code generation presents challenges that extend well beyond those encountered in generating individual functions or files, representing a more complex and underexplored frontier in automated software engineering research.

Recent research has made initial progress in project-level code generation.
Several works have constructed preliminary datasets to support end-to-end project synthesis \cite{liu2025projecteval,li2025prompting,zhang2024experimenting}, and some approaches employ LLMs within pipeline frameworks or multi-agent systems to progressively generate complete project code under structured guidance \cite{hong2023metagpt,qian2024chatdev,zan2024codes}.
However, existing approaches still encounter several limitations:

\ding{172} \textbf{Insufficient Realism and Reliability in Existing Datasets and Metrics}. Current project-level code generation datasets mainly consist of small-scale projects with no more than five files or 500 lines of code, which may fail to capture the complexity of real-world software requirements and project structures.
Most datasets still rely on statistical or similarity-based metrics for evaluation, which may not accurately reflect the actual quality of the generated code.
For example, the SketchBLEU metric \cite{zan2024codes}, an extension of CodeBLEU \cite{ren2020codebleu} for repository-level evaluation, can assign a score of over 50 out of 100 to projects that fail to compile or run.
Therefore, there still lack datasets that reflect real-world project scenarios and support evaluation through executable test cases. 

\ding{173} \textbf{Semantic Gap Between Requirements and Executable Code}. Most existing approaches directly provide requirement documents from software engineering practice to LLMs, which may limit generation performance. Specifically, such requirement documents are primarily written for human understanding and may contain multi-modal elements, such as diagrams, tables, or informal descriptions, which are difficult for LLMs to interpret effectively. Consequently, there is a need for a structured intermediate representation that is both machine-interpretable and human-readable, in order to bridge the semantic gap between natural language requirements and executable code, while enabling systematic quality control throughout the generation process.

\ding{174} \textbf{Hierarchical Contextual Dependencies and Long-Range Context Management Challenges}. Project-level code generation is complicated by both the hierarchical structure of software repositories and the extensive contextual dependencies across generation stages.
Software repositories typically encompass multiple layers, such as directories, files, classes, and functions, which interconnected through intricate relationships of containment, invocation, and inheritance. 
Errors introduced at early stages can propagate through dependent components, amplifying their impact on overall repository quality, whereas existing approaches often conduct assessment and refinement until after the final code has been produced.
Besides, the limited context length of current LLMs constrains their ability to maintain and utilize long-range dependencies throughout the generation process.
Therefore, performing quality control and intermediate evaluation at each generation stage is essential to ensure correctness and consistency, while it is also crucial to consistently retain the most relevant information within the context.

To address these limitations, we first curate and reconstruct a project-level code generation dataset, \ourdata{}, with each task requires generating 12.7 source files and 2,388.6 lines of code on average. 
We select 18 high-quality real-world repositories and augment them with corresponding documentation that reflects user requirements.
Compared to existing datasets \cite{hong2023metagpt, ma2025srlcg, li2025prompting}, \ourdata{} better captures the scale and structural complexity of real-world software development.
Furthermore, we propose \ourmethod{}, a multi-agent project code generation framework that decomposes complex task into manageable sub-tasks via sequential architecture design, skeleton generation, and code filling.
In the architecture design stage, we introduce Semantic Software Architecture Tree (SSAT), a tree-structured, machine-parsable representation of modules, files, and functions with their responsibilities and dependencies, enabling structured and controllable code generation.
The skeleton generation stage produces a scaffold with functions initially filled with pass, which is then expanded into complete project code in the code filling stage.
To address the difficulty of project-level generation, \ourmethod{} incorporates iteration-based refinement guided by judge feedback and a memory-based context management mechanism.
Experimental results show that \ourmethod{} achieves state-of-the-art performance on both small-scale and larger-scale project-level code generation tasks.
For small-scale code generation, \ourmethod{} passes 52 out of 124 test cases on DevBench \cite{li2025prompting}, representing a 57\% improvement over the baseline approaches. 
For larger-scale code generation, \ourmethod{} passes 310 test cases on \ourdata{}, approximately ten times higher than the baseline approaches.

Our contributions can be summarized as follows:
\begin{itemize}
    \item We identify key limitations in project-level code generation, pinpointing the challenges of unrealistic datasets and metrics, semantic misalignment between requirements and executable code, and the difficulty of handling hierarchical dependencies and long-range context during generation.
    \item We build \ourdata{}, a project-level code generation dataset with executable test cases provided for automatic evaluation, which better reflects the scale and structural complexity of real-world software development compared to existing datasets.
    \item We propose \ourmethod{}, a multi-agent three-stage framework for project-level code generation, which incorporates memory-based iterative optimization to progressively refine implementations from user requirements. Within this framework, we introduce the Semantic Software Architecture Tree (SSAT), a structured and semantically rich representation that bridges natural language requirements and source code implementation.
    \item We conduct a comprehensive evaluation of \ourmethod{}, and the results indicate that \ourmethod{} achieves the state-of-the-art performance on both small-scale and larger-scale project-level code generation tasks. Our code and data can be found at \url{https://github.com/whisperzqh/ProjectGen}.
    
\end{itemize}

\section{Related Work}

\subsection{Datasets for Project-level Code Generation}

The project-level code generation task aims to generate a structurally coherent and functionally complete multi-file software project directly from high-level user requirements \cite{li2025prompting}.
Most existing work on code generation focuses on repository-level function generation \cite{zhang2023repocoder, wang2024rlcoder, zhang2024codeagent, ding2023crosscodeeval, wu2024repoformer}. While such tasks do involve code from the entire repository as context, the actual generation is still limited to individual functions. In contrast, project-level code generation requires producing an entire codebase from scratch, encompassing multiple files.
At present, there is a limited availability of datasets specifically designed for project-level code generation, which poses a challenge for developing and evaluating models at this scale.

Some datasets provide only project requirements without corresponding ground-truth code, and the evaluation is based on statistical metrics rather than direct comparison with reference implementations.
For instance, SRDD \cite{qian2024chatdev} comprises 1,200 software task prompts carefully categorized into five main domains: Education, Work, Life, Game, and Creation. Each domain is further subdivided into 40 subcategories, with each subcategory containing 30 unique task prompts. It is evaluated using metrics such as completeness (the percentage of software without any placeholder code snippets), executability (the percentage of software that compiles successfully and can run directly), and consistency (the cosine distance between the semantic embeddings of the textual requirements and the generated software code).
SRLCG dataset \cite{ma2025srlcg} contains 400 samples, each including task descriptions, functional requirements, and technical specifications. The generated code is evaluated in terms of code length, project completeness (assessed via LLMs), and human evaluation. 
CASSD dataset \cite{zhang2024experimenting} consists of 72 tasks encompassing small games, personal websites, and other applications. Evaluation for this dataset is typically based on use case coverage, reflecting whether the generated code satisfies the intended functionality.
SoftwareDev \cite{hong2023metagpt} contains 70 tasks covering Python game development, CRUD application generation, and simple data analysis. Except for statistical evaluation, it further introduces execution-based evaluation, assessing whether the generated code can be successfully executed.

Since evaluation based on statistical metrics or heuristics is often indirect and can be subjective, more recent datasets have started to include ground-truth code, enabling direct comparison between generated code and reference implementations, which allows for more objective and reliable evaluation.
SketchEval \cite{zan2024codes} is a dataset of 19 real-world repositories of varying complexity, which are delicately selected from the latest open-source GitHub projects. Following the idea of CodeBLEU \cite{ren2020codebleu}, it designs SketchBLEU metric to calculate the repository-level similarity between the generated repository with the reference one.

However, evaluating code correctness solely based on similarity to reference implementations has clear limitations. High textual similarity does not necessarily indicate high code quality, as functionally correct or semantically equivalent implementations may differ substantially in structure or style.
To enable a more fine-grained assessment of functional correctness, DevBench \cite{li2025prompting} integrates test cases into its evaluation framework, providing a more reliable and objective assessment. It contains 22 repositories across four programming languages (Python, C/C++, Java, JavaScript), spanning various domains such as machine learning, web services, and command-line utilities. However, the scale of these repositories remains relatively small, making it difficult to accurately reflect the complexity and structure of real-world software development scenarios.

\begin{table}[t]
    \centering
    \setlength{\abovecaptionskip}{0.1cm}
    \caption{Comparison of existing project-level code generation datasets. \textit{\#FILE}: average number of code files contained in ground-truth implementation of each task; \textit{\#LOC}: average lines of code; \textit{Complexity}: average value of cyclomatic complexity; \textit{\#Tests}: average number of unit tests; \textit{Test Coverage}: average coverage of unit tests in each task.}
    \resizebox{\textwidth}{!}{
    \begin{tabular}{lcccccccc}
        \toprule
        \textbf{Datasets} & \textbf{Test Type} & \textbf{Automated-Eval} & \textbf{\#Tasks} & \textbf{\#FILE} & \textbf{\#LOC} & \textbf{Complexity} & \textbf{\#Tests} & \textbf{Test Coverage}\\
        \midrule
        SRDD \cite{qian2024chatdev} & statistics & \cmark & 1,200 & N/A & N/A & N/A & N/A & N/A\\
        SRLCG \cite{ma2025srlcg} & statistics & \xmark & 400 & N/A & N/A & N/A & N/A & N/A\\
        CASSD \cite{zhang2024experimenting} & use cases & \xmark & 72 & N/A & N/A & N/A & N/A & N/A\\
        SoftwareDev \cite{hong2023metagpt} & execution & \xmark & 70 & N/A & N/A & N/A & N/A & N/A\\
        SketchEval \cite{zan2024codes} & similarity & \cmark & 19 & 10.2 & 1,953.1 & 2.93 & N/A & N/A\\
        DevBench$_{Python}$ \cite{li2025prompting} & test cases & \cmark & 10 & 2.2 & 276 & 3.51 & 12.4 & 91.8\%\\
        
        \rowcolor{gray!20}
        \ourdata & test cases & \cmark & 18 & 12.7 & 2,388.6 & 3.03 & 186 & 90.7\% \\
        \bottomrule
    \end{tabular}
    }
    \label{tab:benchmark_comparison}
    \vspace{-0.3cm}
\end{table}

Table \ref{tab:benchmark_comparison} shows the comparison between these project-level code generation datasets. From the results we can observe that existing datasets exhibit several limitations in terms of scale, evaluation methodology, and task complexity. 
The majority of existing datasets do not provide ground-truth implementations, preventing effective assessment of task difficulty and reliable evaluation of the code generated by LLMs.
Only DevBench \cite{li2025prompting} supports automated test-case-based evaluation, but the scale of its repositories is insufficient to reflect real-world software development scenarios.
As a result, existing datasets continue to fall short in reflecting real-world project settings while providing support for evaluation based on executable test cases.

\subsection{Approaches for Project-level Code Generation}

Project-level code generation requires holistic reasoning over architecture design, module partitioning, and dependency management to ensure both local correctness and global consistency. Due to its complex design constraints and multi-dimensional objectives, it remains one of the most challenging problems in code generation research. Recent studies suggest that such task may exceed the capabilities of a single LLM, and that iterative generation with feedback or the collaboration of multiple specialized LLMs is essential for effective task decomposition, reasoning, and refinement throughout the development process.

CodeS \cite{zan2024codes} proposes a simple yet effective framework for the project-level code generation task by decomposing project-level code generation into multiple sub-tasks through a multi-layer sketching process. It consists of three core modules: RepoSketcher, which generates the repository’s directory structure from requirements; FileSketcher, which produces file-level sketches based on the structure; and SketchFiller, which completes the detailed implementation of each function within the generated sketches.
Nevertheless, its generation process follows a one-shot paradigm, lacking intermediate validation or iterative refinement inside or across stages. Consequently, inaccuracies introduced in earlier phases, such as repository structuring or file sketching, tend to propagate throughout subsequent stages, leading to cumulative errors and a decline in the overall coherence and quality of the generated project.
Moreover, SRLCG \cite{ma2025srlcg} introduces a self-rectified framework for generating complete multi-file software projects from a single prompt. It employs a multidimensional chain-of-thought mechanism together with self-rectification, enabling LLMs to produce more accurate and robust code files. These files are then integrated into a coherent project through a dynamic backtracking algorithm, to improve structural consistency and correctness across the entire codebase.

Recently, researchers have explored agent-based frameworks, leveraging multiple LLM-driven agents to collaboratively tackle project-level code generation.
MetaGPT \cite{hong2023metagpt} adopts a multi-agent collaboration framework based on standard operating procedures, simulating the complex software development process as a division of labor and cooperation among different roles. This design enhances the controllability of task decomposition and information flow. 
In each step agents with specific roles generate solutions by adhering to static instructions predefined by human experts.
It provides a global memory pool to store all collaboration records where each agent can subscribe to or search for the information they require.
ChatDev \cite{qian2024chatdev} is a chat-based software development framework that employs specialized LLM-driven agents to collaboratively perform software engineering tasks. Guided by a structured chat chain and an interactive de-hallucination mechanism, these agents coordinate through multi-turn dialogues to accomplish design, coding, and testing, while maintaining conversational coherence through the integration of short-term and long-term memory.
However, it primarily targets small-scale projects with simple requirements, as it takes a single task description as input. Moreover, its memory mechanism merely retains the most recent information in chronological order, which makes it difficult to ensure that the most relevant context is preserved when the conversation becomes lengthy.

\section{Construction of \ourdata{}}

To better reflect real-world project scenarios and support evaluation through executable test cases, we construct a new project-level code generation dataset, \ourdata{}, which consists of 18 Python repositories covering a wide range of topics.
The construction was carried out by four of the authors, each possessing over three years of Python programming experience and research experience in LLM-based code generation.
In this section, we first introduce the construction progress of \ourdata{}, including repository collection and documentation preparation. We then provide a brief overview of \ourdata{} as well as the comparison with existing datasets.

\subsection{Repository Collection}

To efficiently acquire high-quality Python repositories, we avoided the costly process of manually filtering repositories from scratch on GitHub.
Instead, we utilized existing datasets that were originally developed for related code generation tasks and already contain curated and well-maintained repositories.
In particular, the selected datasets should meet two important requirements: \ding{172} they contain complete Python repositories; \ding{173} they are accompanied by unit tests, enabling an executable evaluation of the correctness and functionality of the code within repository.  
Consequently, these datasets provide a reliable foundation of high-quality repositories that are well aligned with the objectives of our dataset construction.
Specifically, we collected Python repositories from the following three datasets:

\begin{itemize}
    \item \textbf{Commit0} \cite{zhaocommit0}: This is a large-scale benchmark designed to evaluate LLM’s ability to generate fully functional Python code, which consists of 54 Python libraries. Each task provides a specification document and a starter repository with unit tests, and LLMs are required to fill in the missing source code to achieve full test pass rates.
    \item \textbf{DevEval} \cite{li2024deveval} : This is a comprehensive benchmark designed to evaluate the coding capabilities of LLMs in real-world software development scenarios. It includes 1,874
    repository-level function code generation tasks derived from 117 real-world Python repositories with unit tests for evaluation.
    \item \textbf{CoderEval} \cite{yu2024codereval}: This is a repository-level function code generation benchmark, which includes 230 Python generation tasks curated from 43 real-world open-source projects. Each task is accompanied by a self-contained execution environment for automatic functional evaluation.
\end{itemize}

To select suitable repositories from the aforementioned sources, we first collected key statistics for each repository. 
We used the \texttt{radon}\footnote{https://pypi.org/project/radon} library to compute the number of code files, lines of code, and code complexity. We also employed \texttt{pytest}\footnote{https://pypi.org/project/pytest} library to measure the number of unit tests and the coverage.
Based on these statistics, we then established a set of filtering criteria to systematically identify repositories that meet our quality and completeness requirements, which can be seen as follows:

\begin{itemize}
    \item \textbf{Repository size constraint:} To ensure that the repositories reflect project-level code generation while remaining within the generation capability of current LLMs, we select projects whose total lines of code fall within the range of 500 to 10,000.
    \item \textbf{Structural complexity requirement:} To guarantee that each repository exhibits non-trivial architectural design, only projects containing three or more source files are considered.
    \item \textbf{Test coverage threshold:} To ensure that unit tests can effectively assess the quality of generated code, we select repositories with at least 80\% unit test line coverage.
    \item \textbf{Topical diversity consideration:} To enhance the representativeness of the dataset, we also take into account the thematic diversity of repositories, ensuring coverage across a wide range of domains and tasks.
\end{itemize}

Ultimately, a total of 18 repositories were selected for the construction of \ourdata{}, comprising 9 from Commit0, 7 from DevEval, and 2 from CoderEval.

\begin{table}[t]
    \centering
    \setlength{\abovecaptionskip}{0.1cm}
    \caption{Detailed statistics of \ourdata{}. We provide both the value for each repository and the corresponding averages and medians, with the maximum values highlighted using underlines. \textit{\#PRD tokens} stands for tokens of PRD calculated by the DeepSeek tokenizer, and \percent{cov.} represents the unit test line coverage rate.}
    \resizebox{\textwidth}{!}{
    \begin{tabular}{ccccccc}
        \toprule
        \textbf{Repository} & \textbf{\#FILE} & \textbf{\#LOC} & \textbf{Complexity} & \textbf{\#Check Tests \percent{(cov.)}} & \textbf{\#Unit Tests \percent{(cov.)}} & \textbf{\#PRD tokens} \\
        \midrule
        bplustree & 8 & 1,509 & 2.29 & 8 \percent{(82\%)} & 356 \percent{(98\%)} & 1,339\\
        cookiecutter & 18 & 2,805 & 3.42 & 7 \percent{(55\%)} & \uline{375} \percent{(99\%)} & 2,100\\
        csvs-to-sqlite & 3 & 816 & \uline{5.83} & 10 \percent{(81\%)} & 25 \percent{(88\%)} & 1,841\\
        deprecated & 3 & 597 & 4.08 & \uline{26} \percent{(80\%)} & 176 \percent{(95\%)} & 953\\
        djangorestframework-simplejwt & \uline{31} & 2,014 & 2.09 & 8 \percent{(63\%)} & 191 \percent{(93\%)} & 1,614\\
        flask & 24 & \uline{9,314} & 2.71 & 25 \percent{(52\%)} & 482 \percent{(91\%)} & 2,913\\
        imapclient & 17 & 3,531 & 2.81 & 9 \percent{(40\%)} & 267 \percent{(80\%)} & \uline{3,810}\\
        parsel & 5 & 1,128 & 2.60 & 5 \percent{(65\%)} & 250 \percent{(95\%)} & 1,522\\
        portalocker & 9 & 1,958 & 2.84 & 10 \percent{(58\%)} & 71 \percent{(94\%)} & 1,990\\
        pyjwt & 12 & 2,690 & 3.01 & 10 \percent{(53\%)} & 294 \percent{(94\%)} & 382\\
        python-hl7 & 11 & 2,434 & 2.98 & 10 \percent{(56\%)} & 100 \percent{(87\%)} & 2,292\\
        rsa & 14 & 2,949 & 2.40 & 6 \percent{(73\%)} & 100 \percent{(87\%)} & 3,318\\
        simpy & 12 & 2,184 & 2.01 & 7 \percent{(60\%)} & 149 \percent{(90\%)} & 2,147\\
        tinydb & 10 & 2,170 & 1.76 & 10 \percent{(58\%)} & 204 \percent{(95\%)} & 947\\
        trailscraper & 13 & 890 & 2.01 & 4 \percent{(65\%)} & 93 \percent{(92\%)} & 3,415\\
        voluptuous & 7 & 3,100 & 2.55 & 11 \percent{(55\%)} & 161 \percent{(90\%)} & 1,221\\
        xmnlp & 24 & 1,504 & 3.47 & 8 \percent{(65\%)} & 23 \percent{(81\%)} & 3,105\\
        zxcvbn & 8 & 1,402 & 5.69 & 6 \percent{(81\%)} & 31 \percent{(84\%)} & 2,399\\
        \midrule
        \rowcolor{gray!20}
        \textbf{Avg.} & \textbf{12.7} & \textbf{2388.6} & \textbf{3.03} & \textbf{10 (63.4\%)} & \textbf{186 (90.7\%)} & \textbf{2,067}\\
        \rowcolor{gray!20}
        \textbf{Mid.} & \textbf{11.5} & \textbf{2092} & \textbf{2.76} & \textbf{8.5 (61.5\%)} & \textbf{168.5 (91.5\%)} & \textbf{2,100}\\
         \bottomrule
    \end{tabular}
    }
    \label{tab:dataset_construction}
\end{table}

\subsection{Documentation Preparation} \label{sec:document_preparation}

To better simulate real-world software development practices, we systematically reverse-engineered documentation from selected repositories. 
Following \citet{li2025prompting}, we reconstructed the product requirement document (PRD), UML diagrams, and architectural design specifications for each repository.
Furthermore, we categorized the test cases provided within the repositories into \textit{check tests} and \textit{unit tests}, with the details presented below.

\subsubsection{Construction Principles}
To ensure a more standardized construction process and maintain consistency in structure and style across files generated by different data curators, we define a unified set of guidelines specifying the construction rules for each kind of documentation.

\textbf{PRD:} PRD provides detailed descriptions of a software system’s functional and non-functional requirements, guiding subsequent design and development.
The document was produced through a hybrid approach, combining automated processes with manual refinement.
Curators are required to use DeepWiki\footnote{https://deepwiki.com} to automatically analyze the repository and extract its key functional elements.
DeepWiki is an AI-powered tool developed by Cognition AI that transforms public GitHub repository into an interactive and structured knowledge base, allowing users to access automatically generated documentation and AI-driven Q\&A capabilities.
Based on the documentation generated by DeepWiki and the original README files in the repository, curators manually refined them into a PRD encompassing the introduction, goals, features and functionalities, usage, requirements, \textit{etc}.

\textbf{UML diagrams:} For UML diagrams, curators are required to use Pyreverse\footnote{https://pylint.readthedocs.io/en/latest/additional\_tools/pyreverse} for automatically generation. Pyreverse is a tool included in the Pylint package that analyzes Python code and produces UML class and package diagrams, helping developers visualize the structure and relationships of the repository.
With the code repository as input, Pyreverse can automatically generate the corresponding class and package diagrams in mermaid format.

\textbf{Architecture design:} The architecture design document mainly comprises two parts. The first part depicts the directory tree of the repository, including only the source files and excluding test files and other non-essential files, which is generated directly by using Linux \texttt{tree} command. 
The second part provides concise descriptions for each source file, accompanied by summaries of the classes and functions they contain, aiming to clarify the functional responsibilities and modular decomposition of the repository. 
When classes or functions already include comments in the source code, data curators directly adopted these existing annotations. For elements lacking documentation, curators employed LLMs to generate appropriate descriptions, which were subsequently manually reviewed to ensure both accuracy and conciseness.

\textbf{Test cases:} We categorize the test cases into \textit{check tests} and \textit{unit tests}.
Check tests serve to provide initial verification during the code generation process, allowing LLMs to iteratively refine the code based on test feedback. Unit tests are executed upon completion of code generation to evaluate the overall quality and functional correctness of the generated projects.
To reduce the risk of data leakage, the check tests and unit tests are ensured to be entirely disjoint sets.
During the construction process, data curators manually split the original test cases provided in each repository. 
Initially, all repository-provided tests were added into the unit test set.
Subsequently, curators created the check tests by selecting the highest-level unit tests and manually modifying their inputs and outputs
Tests that could not be feasibly adapted without compromising their validity, including those involving external dependencies or complex data constraints, were incorporated directly into the check test set and removed from the unit test set.
Consequently, unit test coverage is maintained above 80\%, and the number of check tests remains below ten for most repositories. This design reflects real-world development practices, where programmers typically provide only approximate test plans prior to full code implementation rather than comprehensive validation suites.

\subsubsection{Construction Pipeline}
Among the four data curators, one served as the team lead, responsible for overall coordination, quality review, and making final arbitration in cases of disagreement.
The remaining three curators each handled six of the 18 repositories independently, with allocation balanced by complexity to ensure manageable workloads and consistent annotation quality.

All curators strictly followed the standard software development workflow to systematically analyze each repository: PRD $\rightarrow$ UML diagrams $\rightarrow$ architectural design $\rightarrow$ test cases.
We also prepared complete environment dependencies and configuration files for each repository.
This structured approach enable the curators to gain a comprehensive understanding of each repository's context and design rationale. 
To ensure high-quality and consistent construction, we followed a three-stage protocol. Initially, individual curators performed the first round of construction. This was followed by a peer cross-review, in which potential discrepancies were identified and corrected through mutual validation. Finally, the team lead conducted a comprehensive review to resolve any remaining ambiguities or disagreements.

\subsection{Overview of \ourdata{}}

Table \ref{tab:dataset_construction} presents the detailed statistics of the 18 repositories included in \ourdata{}. On average, each repository contains 12.7 source files and 2,388.6 lines of code, with 10 check tests achieving 63.4\% coverage and 186 unit tests achieving 90.7\% coverage. The median values are close to these averages, indicating that the dataset maintains a balanced distribution across repositories.
In terms of requirements, the average number of tokens in the PRD documents is 2,067, reflecting the complexity and richness of the specifications provided as input for project-level code generation.

Furthermore, as shown in Table \ref{tab:benchmark_comparison}, \ourdata{} demonstrates clear advantages in terms of scale, completeness, and evaluation reliability. It provides fully runnable projects equipped with comprehensive unit and check tests, supporting automated and objective evaluation of code generated by LLMs. \ourdata{} also captures realistic project structures, featuring a moderate level of architectural complexity and diverse software functionalities that better reflects real-world development scenarios. 
In addition, the abundant test cases and consistently high test coverage ensure rigorous validation of both correctness and completeness. Collectively, these characteristics establish \ourdata{} as a comprehensive, challenging, and reliable benchmark for advancing research on project-level code generation.

\section{Approach}
To advance the capability of project-level code generation, we propose \ourmethod{}, a three-stage framework that leverages coordinated multi-agent collaboration to progressively synthesize complete software projects.
We first introduce the Semantic Software Architecture Tree, a structured and semantically enriched representation that extracts and organizes essential information from user requirements, to effectively narrow the gap between user-facing documentation and executable source code.
Based on this, \ourmethod{} reformulates project-level code generation as a progressive three-stage process comprising architectural design, skeleton generation, and code filling. Each stage is collaboratively executed by a generation agent and a judge agent, aiming at alleviating the cascading effect of early-stage errors by systematic progression and validation.

\subsection{Semantic Software Architecture Tree}

In practical software engineering, a substantial semantic and representational gap often exists between high-level requirement specifications and executable source code. Architectural design plays a critical intermediary role in bridging this gap by organizing system-level intent into implementable structures.
Motivated by this observation, we introduce the Semantic Software Architecture Tree (SSAT), a structured and semantically grounded representation that systematically distills salient information from user requirements.
Unlike conventional textual descriptions, SSAT adopts a tree-structured, machine-parsable format that explicitly encodes modules, files, and functions together with their responsibilities. Such a representation not only preserves the design rationale in a human-readable form but also aligns naturally with the compositional structure of source code, enabling LLMs to interpret architectural intent and progressively generate implementation-level artifacts.
Consequently, SSAT serves as a bidirectional component to enhance both interpretability and controllability in project-level code generation.

\begin{figure}[t]
  \centering
  \includegraphics[width=\linewidth]{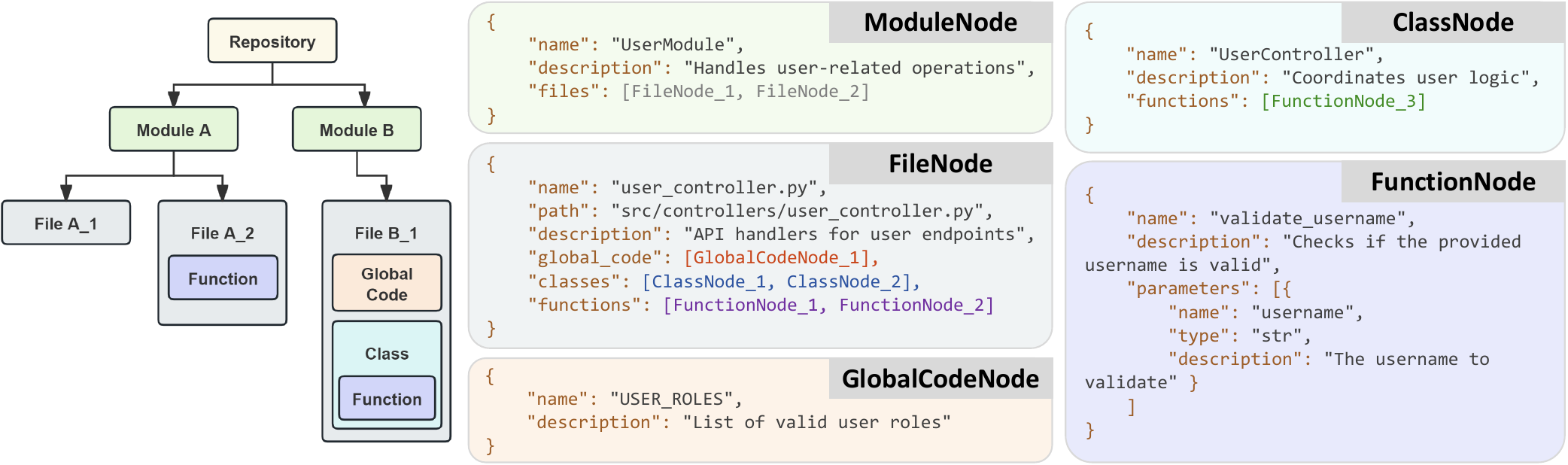}
  \caption{Detailed illustration of Semantic Software Architecture Tree. The left part shows the hierarchical organization of a repository, and the right part provides detailed examples of the elements contained in each type of node. }
  \label{fig:ssat}
  \vspace{-0.3cm}
\end{figure}

Figure \ref{fig:ssat} provides a detailed illustration of the structure of SSAT, which comprises five types of nodes: \textsc{ModuleNode}, \textsc{FileNode}, \textsc{GlobalCodeNode}, \textsc{ClassNode}, and \textsc{FunctionNode}.
Each type of node contains two common attributes: \textit{name} and \textit{description}. 
As files are typically organized directly under the repository root in practical code repositories, the \textsc{ModuleNode} serves as a logical abstraction rather than a physically existing entity, to facilitate a clearer understanding of the relationships between the overall architecture and individual files.
Thus, except for the \textsc{ModuleNode}, the \textit{name} attribute of each node corresponds to its actual name in the code repository to be generated.
Below, we provide a detailed description of each node type, along with their distinct attributes.

\noindent\textsc{\textbf{ModuleNode}}: Each represents a logical unit with a relatively independent functionality.
\begin{itemize}[nosep]
    \item \textit{files}: the set of all \textsc{FileNode} contained within the module.
\end{itemize}

\noindent\textsc{\textbf{FileNode}}: Each corresponds to a file in the code repository. Based on empirical observation, the contents of a file can be categorized into defined classes, functions, and global code outside of these definitions.
\begin{itemize}[nosep]
    \item \textit{path}: the location of the file relative to the root directory of the code repository to be generated.
    \item \textit{global\_code}: the set of all \textsc{GlobalCodeNode} contained within the file.
    \item \textit{classes}: the set of all \textsc{ClassNode} contained within the file.
    \item \textit{functions}: the set of all \textsc{FunctionNode} contained within the file.
\end{itemize}

\noindent\textsc{\textbf{ClassNode}}: Each represents a class defined within a file in the code repository. 
\begin{itemize}[nosep]
    \item \textit{functions}: the set of all \textsc{FunctionNode} contained within the class.
\end{itemize}

\noindent\textsc{\textbf{FunctionNode}}: Each represents a function defined either at the file level (global function) or within a class.
\begin{itemize}[nosep]
    \item \textit{parameters}: the set of input parameters for the function. Each parameter is further characterized by three elements: \textit{name}, \textit{type}, and \textit{description}, capturing the parameter’s identifier, its data type, and a brief explanation of its role or purpose within the function.
\end{itemize}

As shown in the left part of Figure \ref{fig:ssat}, typically, a repository contains one or more \textsc{ModuleNode}, and each \textsc{ModuleNode} groups related source units and acts as a logical subsystem. \textsc{ModuleNode} contain \textsc{FileNode}. Each \textsc{FileNode} may include one or more of the following child element types: \textsc{GlobalCodeNode}, \textsc{ClassNode} and \textsc{FunctionNode}. Arrows indicate parent–child containment and the tree topology that enables traversal from repository down to functions and classes. This hierarchical layout emphasizes modular decomposition and the natural mapping from high-level design to concrete source artifacts.

In summary, SSAT is designed to facilitate LLM comprehension and code generation through three complementary aspects.
\ding{172} From the perspective of format, SSAT employs a hierarchical, JSON-like tree with clearly labeled nodes and explicit parent–child relationships, which is easily tokenized and enables multi-level contextual encoding while reducing textual ambiguity. 
\ding{173} From the perspective of content, SSAT captures structural organization along with module responsibilities, interface definitions, and dependencies, distilling implicit design intentions into actionable intermediate representations.
\ding{174} From the perspective of LLM-guidance, the hierarchical organization of SSAT corresponds directly to the layered structure of source code, thereby facilitating systematic and top-down code generation.
Thus, these properties allow SSAT to bridge human-authored documentation and code, guiding LLMs to better interpret requirements, maintain consistency across components, and produce code that faithfully reflects the overall design.

\begin{figure}[t]
  \centering
  \includegraphics[width=\linewidth]{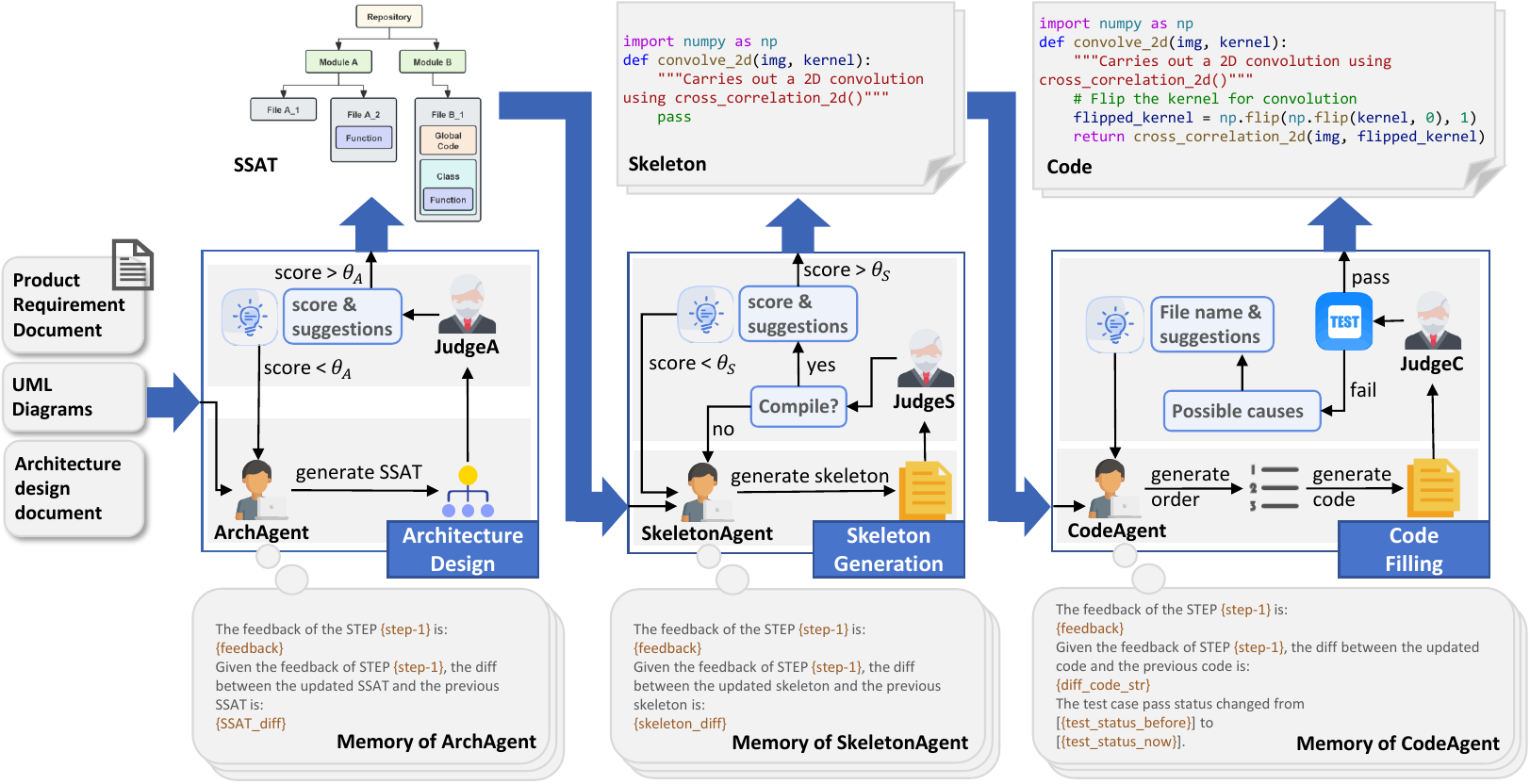}
  \caption{The workflow of \ourmethod{}. It takes user requirements as input and follows three sequential stages: architecture design, skeleton generation, and code filling, where the output of each stage serves as the input for the subsequent stage. }
  \label{fig:architecture}
  \vspace{-0.3cm}
\end{figure}

\subsection{Three-Phase Framework for Project-Level Code Generation}

As illustrated in Figure \ref{fig:architecture}, \ourmethod{} organizes the project-level code generation process into three sequential phases: architecture design, skeleton generation, and code filling.
To execute this workflow, we deploy multiple specialized agents with distinct roles, collaboratively coordinating to accomplish tasks both within individual phases and across the entire process.
In this section, we first present the overall workflow of \ourmethod{}, including the objectives, inputs, and outputs of each phase. The detailed internal processes of each phase will be elaborated in the following section.

In the \textbf{architecture design phase}, \ourmethod{} is responsible for transforming the provided software specifications into an SSAT.
To accomplish this, \ourmethod{} is first provided with a concise description of the generation task and an explicit introduction to the concept and structure of the SSAT, ensuring it understands both the goal and the expected representation format.
In real-world software development, code implementation is typically grounded in a set of well-defined artifacts, including requirement specifications, architectural designs, and preliminary design diagrams.
These documents collectively guide developers from conceptual intent to structural realization, providing complementary perspectives on what to build, how the system is organized, and how components interact.
Motivated by this observation, and following \cite{li2025prompting}, \ourmethod{} adopts three types of inputs that collectively describe the target repository from complementary perspectives:
\begin{itemize}
    \item \textit{Product Requirement Document (PRD):} a natural-language specification detailing the repository’s functional and non-functional requirements, component responsibilities, and user interactions.
    \item \textit{UML Diagrams:} which include class and/or sequence diagrams in mermaid format, providing structural and behavioral information about the system. These diagrams describe class relationships, defined methods with parameter signatures, and dynamic interaction flows such as call orders and parameter passing across modules.
    \item \textit{Architecture Design Document:} a high-level specification of the repository’s organizational structure, defining the directory hierarchy, file responsibilities, and module boundaries.
\end{itemize}
These documentations are sequentially presented in the prompt, followed by detailed extraction instructions and a formal specification of SSAT format. 
Based on this, \ourmethod{} synthesizes the architectural information into an SSAT.

In the \textbf{skeleton generation phase}, \ourmethod{} aims at iteratively generating the structural skeleton of each source file based on the SSAT derived from the preceding architecture design phase.
\ourmethod{} is first presented with an explicit task description along with an overview of the SSAT’s hierarchical organization. Subsequently, \ourmethod{} is provided with the SSAT representation of the target file, detailing modules, classes, functions, and associated metadata.
Leveraging this information, \ourmethod{} produces a full code skeleton encompassing import statements, global definitions, class declarations, and function signatures with placeholder \texttt{pass} bodies. Each function signature strictly adheres to the parameter and default value constraints defined in the SSAT, and concise descriptive comments are inserted immediately beneath the signatures.
Under ideal conditions, the generated skeletons are expected to be fully syntactically valid and directly compilable, providing a reliable scaffold for the subsequent implementation phase.

In the \textbf{code filling phase}, \ourmethod{} focuses on transitions from skeleton to concrete implementation.
\ourmethod{} receives the skeleton of the target file along with the contextual information from previously generated files to ensure cross-file consistency.
Its task is to fill in function bodies while strictly preserving the original structure of classes and functions as defined in the skeleton. Descriptive comments from the skeleton are retained, and all naming conventions, imports, and shared components are maintained in accordance with the repository context. The ideal output is a complete and syntactically valid code file, that can be directly integrated into the overall project.
By systematically applying this procedure to each file defined in the skeleton, \ourmethod{} ultimately produces the fully implemented project.

\subsection{Iterative Code Optimization with Long-Context Management}

Given the inherent complexity of project-level code generation and the significant semantic gap between high-level requirements and executable code, achieving satisfactory code repository through a single-pass sequential generation is highly challenging. 
Therefore, \ourmethod{} adopts a multi-agent cooperative paradigm that enables both content generation and quality assurance within each phase, thereby mitigating the propagation of errors from early stages and ensuring the overall integrity and correctness of the generated project.
Specifically, each phase involves a pair of agents: a \textit{generation agent} responsible for producing outputs, and a corresponding \textit{judging agent} tasked with evaluating their quality and ensuring compliance with predefined standards before proceeding to the next stage.
Concretely, \textbf{\textsc{ArchAgent}}, \textbf{\textsc{SkeletonAgent}}, and \textbf{\textsc{CodeAgent}} undertake architecture design, skeleton generation, and code filling, respectively.
\textbf{\textsc{JudgeA}}, \textbf{\textsc{JudgeS}}, and \textbf{\textsc{JudgeC}} rigorously assess the generated results at each stage.
Each generation agent leverages a memory mechanism to access relevant contextual information from previous iterations, facilitating iterative refinement. To maintain the independence and objectivity of evaluations, the judging agents do not utilize memory.

The workflow of \ourmethod{} can be formally described as a sequential, iterative multi-agent process with integrated quality control.
Let $\mathcal{G} = \{\text{\textsc{ArchAgent}}, \text{\textsc{SkeletonAgent}}, \text{\textsc{CodeAgent}}\}$ denote the set of generation agents responsible for each phase, and $\mathcal{J} = \{\text{\textsc{JudgeA}}, \text{\textsc{JudgeS}}, \text{\textsc{JudgeC}}\}$ the corresponding judging agents.
At iteration $t$, the generation agent $G_t \in \mathcal{G}$ of the active phase produces output $O(G_t)$ that is assessed by corresponding $J_{G_t} \in \mathcal{J}$, which returns a binary score indicating whether the output meets the required quality, where 0 indicates unsatisfactory output requiring revision, and 1 indicates satisfactory output allowing transition. The phase update is defined as:
\begin{equation}
    \Phi(G_t) =
    \begin{cases}
    G_t, & \text{if } J_{G_t}(G_t) = 0 \\
    next(G_t), & \text{if } J_{G_t}(G_t) = 1 \text{ or } iter(G_t) \geq N_{\text{max}}
    \end{cases}
\end{equation}
where $iter(G_t)$ denotes the current iteration count within $G_t$, and $N_{\text{max}}$ is the predefined upper bound on the number of iterations for each phase. The next-phase mapping $next(G_t)$ is defined as: 
\begin{equation}
    \operatorname{next}(G_t) =
    \begin{cases}
    \text{\textsc{SkeletonAgent}}, & \text{if } G_t = \text{\textsc{ArchAgent}} \\
    \text{\textsc{CodeAgent}}, & \text{if } G_t = \text{\textsc{SkeletonAgent}} \\
    \text{\textsc{End}}, & \text{if } G_t = \text{\textsc{CodeAgent}}
    \end{cases}
\end{equation}
The workflow initiates with $G_0 = \text{\textsc{ArchAgent}}$ and iteratively updates until $G_t = \text{\textsc{End}}$. 
The detailed procedures for each phase are presented below.

\subsubsection{Architecture Design}

In the architecture design phase, \textsc{ArchAgent} first generates the SSAT and submits to \textsc{JudgeA} for evaluation.
Following the LLM-as-judge paradigm \cite{gu2024survey, li2024llms}, \textsc{JudgeA} is instructed to assess the SSAT from the following perspectives:
\begin{itemize}
    \item \textit{Requirement Coverage:} Does the architecture cover all the functional modules mentioned in the requirements?  
    \item \textit{Consistency with Provided Information:} Does the architecture faithfully follow the directory structure, file names, and function names explicitly given in the PRD, UML diagrams, and Architecture Design Document?
    \item \textit{Interface Consistency:} Are the interface names clear, unambiguous, and free from redundancy?  
    \item \textit{Dependency Relations:} Are there any circular dependencies? Does the dependency structure follow common layered architecture principles? 
\end{itemize}
\textsc{JudgeA} evaluates the generated SSAT across the four dimensions, assigning a score from 0 to 10 for each criterion accompanied by detailed justifications.
An overall score is also required to represent the comprehensive quality.
If the overall score exceeds a predefined threshold $\theta_A$, the phase output progresses to the next stage; otherwise, the feedback provided by \textsc{JudgeA} is returned to \textsc{ArchAgent}, which refines the SSAT iteratively according to the evaluation results.

After each refinement, \ourmethod{} records both the \textsc{JudgeA} feedback and the differences between the previous and updated SSAT. This information is then distilled into a concise summary and is subsequently stored as part of \textsc{ArchAgent}'s \textit{memory}.
This mechanism allows \textsc{ArchAgent} to retain a historical context of past refinements, effectively learning from previous evaluation outcomes. To preserve relevant context while preventing the prompt from becoming excessively long, only the $\gamma_A$ summaries that are most semantically similar to the current \textsc{JudgeA} feedback are selected and included in the prompt for subsequent iterations. Therefore, \ourmethod{} balances the need for historical awareness with computational efficiency, enabling informed and targeted SSAT refinements over multiple iterative cycles.

\subsubsection{Skeleton Generation}
In the skeleton generation phase, \textsc{SkeletonAgent} first produces an initial skeleton based on the SSAT, and then submits it to \textsc{JudgeS} for evaluation.
\textsc{JudgeS} performs a two-stage assessment. It first attempts to compile the skeleton. If the compilation fails, the resulting error diagnostics are returned as feedback to \textsc{SkeletonAgent} for revision. If the skeleton successfully compiles, \textsc{JudgeS} then applies the LLM-as-judge paradigm to perform a higher-level appraisal along two complementary dimensions:
\begin{itemize}
    \item \textit{Directory Structure Matching:} Does the skeleton's directory and file hierarchy match the architecture specification? Are there missing or extra files/directories? Is the nesting consistent with the design?
    \item \textit{Interface \& Call Relationship Matching:} Do the classes and functions (including names, parameters, and default values) align with the architecture definition? Are all expected interfaces present? Are there inconsistencies or omissions?
\end{itemize}
For each dimension, \textsc{JudgeS} produces a numerical score from 0 to 10 and a detailed rationale, as well as an overall score.
If the overall score falls below the threshold $\theta_S$, the feedback is returned to \textsc{SkeletonAgent} for iterative refinement.

Similar to the architecture design stage, the feedback from \textsc{JudgeS} is recorded and combined with the differences between the previous and updated skeleton to form a structured \textit{memory}. During subsequent iterations, only the $\gamma_S$ most relevant pieces of memory are selected and incorporated into the prompt, providing context that guides \textsc{SkeletonAgent} in making informed refinements.

\subsubsection{Code Filling}
In the code filling phase, \textsc{CodeAgent} first determines the generation order of source files.
To establish this order, all valid code files are analyzed to extract their import statements, upon which a topological sorting is performed. This ensures that each file is generated only after the files it depends on, thereby minimizing interface omissions and invocation inconsistencies.

Once the initial version of the project code is generated, it is submitted to \textsc{JudgeC} for evaluation. \textsc{JudgeC} first verifies whether the generated code can successfully pass the predefined check tests.
As discussed in Section \ref{sec:document_preparation}, check tests are used for preliminary validation during the code generation process, which are distinct from test cases.
If the generated code fails to pass all the check tests, \textsc{JudgeC} analyzes the error logs produced during the test execution, identifies the possible causes of the failures, and provides the names of the code files that require modifications along with suggested corrective actions.
This information is then returned to \textsc{CodeAgent} as feedback, guiding it to iteratively revise and improve the existing code.

During iterative code modifications, similar issues may arise across successive versions. To guide \textsc{CodeAgent} using past corrections, \ourmethod{} maintains a \textit{memory} consisting of \textsc{JudgeC} feedback, the differences between pre- and post-modification code, and the corresponding changes in check test pass results. For each new iteration, the $\gamma_C$ pieces of this memory that most closely match the current \textsc{JudgeC} feedback and do not reduce the check test pass rate are retrieved and incorporated into the prompt as contextual information. This approach allows \textsc{CodeAgent} to leverage historical experience to perform targeted refinements while ensuring functional correctness.

\section{Evaluation}

To evaluate the effectiveness of \ourmethod{}, we address the following research questions:
\begin{center}\small
\begin{tcolorbox}[colback=gray!10,
                  colframe=black,
                  arc=1mm, auto outer arc, breakable,
                  boxrule=0.5pt,
                 ]
\begin{itemize}[leftmargin=*]
    \item \textbf{RQ1: Performance on Small-Scale Projects.} How does \ourmethod{} compare with state-of-the-art baseline approaches when generating code for small-scale repositories characterized by a limited number of files and lines of code?
    \item \textbf{RQ2: Performance on Larger-Scale Projects.} How effectively can \ourmethod{} generate code for larger-scale projects comprising thousands of lines of code?
    \item \textbf{RQ3: Impact of Proposed Components.} What is the impact of each individual proposed component of \ourmethod{} on its overall code generation performance?
\end{itemize}
\end{tcolorbox}
\end{center}

\subsection{Experimental Setup}

\subsubsection{Baseline approaches}
We compare \ourmethod{} with three state-of-the-art project-level code generation approaches. All of these approaches take user requirements as input and directly generate complete project-level code as output. Detailed information can be seen as follows:
\begin{itemize}
    \item \textbf{CodeS} \cite{zan2024codes}: It introduces a hierarchical sketching framework that decomposes project-level code generation into multi-granularity sub-tasks through a structured sketching and filling process.
    \item \textbf{MetaGPT} \cite{hong2023metagpt}: It models software development as a multi-agent collaboration workflow guided by predefined standard operating procedures, enabling controllable task decomposition and information exchange among specialized agents.
    \item \textbf{ChatDev} \cite{qian2024chatdev}: It employs a chat-based multi-agent framework where role-specific LLM agents collaborate through structured dialogues to perform software design, coding, and testing under a conversational memory mechanism.
\end{itemize}

\subsubsection{Datasets} 
To support automated evaluation based on test cases, we selected two datasets for experiments, DevBench \cite{li2025prompting} and \ourdata{}.
Specifically, we used the Python subset of DevBench (DevBench$_{Python}$), which consists of 10 tasks with an average of 2.2 files and 276 lines of code per task.
In addition, our constructed dataset \ourdata{} contains 18 tasks, with an average of 12.7 files and 2,388.6 lines of code, representing more complex and realistic projects.
Both datasets include a PRD, UML diagrams, and an architectural design specification for each task.
We follow a similar approach in Section \ref{sec:document_preparation} to create check tests for DevBench, generating an average of 4 tests per task with approximately 70\% coverage.
Given the substantial disparity in scale between the two datasets, we employ DevBench$_{Python}$ for evaluation in RQ1 and utilize \ourdata{} for analysis in RQ2.

\subsubsection{Metrics}
We assess the code generation performance of each approach using the \textbf{number of passed unit tests (\#Pass)} as the primary metric.
Specifically, we record the number of successfully passed unit tests for each task. To obtain the overall performance, following \citet{li2025prompting}, we calculate a weighted average of the passed test counts across all tasks, where each task’s weight corresponds to its total lines of code, thereby partially reflecting the impact of project scale and complexity.

In addition, we employ \textbf{SketchBLEU} \cite{zan2024codes} as a complementary evaluation metric, providing an additional perspective on the structural similarity of the generated code.
It is an extension of the function-level metric CodeBLEU \cite{ren2020codebleu} to the repository level, which can be calculated by:
\begin{equation}
    SketchBLEU=BLEU+BLEU_{weight}+Match_{struc}+Match_{df}
\end{equation}
where $BLEU$ and $BLEU_{weight}$ correspond to token-level evaluations similar to traditional BLEU metrics, $Match_{struc}$ assesses structural similarity based on the abstract syntax tree, and $Match_{df}$ evaluates the similarity of data-flow relations between the generated and reference code.

\subsection{Implementation Details}

For \ourmethod{}, we implement the multi-agent system using LangChain\footnote{https://www.langchain.com}, which is a framework designed to facilitate the development of applications using LLMs.
We use \texttt{DeepDiff}\footnote{https://deepdiff.readthedocs.io} library to capture the differences between the previous and updated SSAT and skeleton, and the \texttt{difflib}\footnote{https://docs.python.org/library/difflib} library to compute code-level differences. When retrieving relevant context from the memory of generation agents, we employ the BM25 \cite{robertson2009bm25} algorithm to measure similarity between feedback texts.
For the hyperparameter settings of experiments on DevBench, based on preliminary experiments, the maximum number of iterations is set to 3, 3 and 5 for the architecture design, skeleton generation, and code filling stages, respectively.
The acceptance thresholds of \textsc{JudgeA} and \textsc{JudgeS} are both set to $\theta_A=\theta_S=8$, while the upper limits on the number of context entries retrieved by the generation agents are set to $\gamma_A = \gamma_S = \gamma_C = 2$.
For experiments on \ourdata{}, we set the maximum number of iterations for code filing stage to 10, with other hyperparameter settings the same as DevBench.
For baseline approaches, to ensure a fair comparison, we provide the PRD, UML diagrams, and architectural design document as inputs, thereby granting them access to the same comprehensive information sources.

We conduct experiments using two backbone LLMs, DeepSeek-V3 \cite{liu2024deepseekv3} and GPT-4o \cite{hurst2024gpt}, both of which have demonstrated strong performance in code understanding and generation tasks.
To ensure the consistency and reproducibility of results, all experiments are conducted with \texttt{temperature=0} and \texttt{top\_p=1.0}. The maximum generation length is set to the upper limit supported by each LLM: 8k tokens for DeepSeek‑V3 and 16k tokens for GPT‑4o.

\section{Results}

\subsection{RQ1: Performance on Small-Scale Projects}

\begin{table}[t]
    \centering
    \setlength{\abovecaptionskip}{0.1cm}
    \caption{Comparison of \textbf{\#Pass} metric between \ourmethod{} and baseline approaches on DevBench, with all non-zero results highlighted with a blue background. We also report the SUM number and AVG$_{weighted}$ which is weighted by the \#LOC of each task. }
    \resizebox{\textwidth}{!}{
    \begin{tabular}{ccc|cccc|cccc}
         \toprule
          \multirow{2}{*}{\textbf{Task}} & \multirow{2}{*}{\textbf{\#LOC}} & \multirow{2}{*}{\textbf{\#Tests}} & \multicolumn{4}{c|}{\textbf{DeepSeek-V3}} & \multicolumn{4}{c}{\textbf{GPT-4o}} \\
          & & & \textbf{ChatDev}& \textbf{CodeS} & \textbf{MetaGPT} & \textbf{\ourmethod{}} & \textbf{ChatDev}& \textbf{CodeS} & \textbf{MetaGPT} & \textbf{\ourmethod{}} \\
         \midrule
         ArXiv\_digest & 198 & 38 & 0 & \cellcolor{c4!10} 24 & \cellcolor{c4!10} 8 & \cellcolor{c4!10} \textbf{25} & 0 & \cellcolor{c4!10} 10 & \cellcolor{c4!10} 23 & \cellcolor{c4!10} \textbf{26}\\
         chakin & 162 & 1 & 0 & 0 & 0 & 0 & 0 & 0 & 0 & 0\\
         geotext & 470 & 4 & 0 & 0 & \cellcolor{c4!10} \textbf{2} & 0 & 0 & 0 & 0 & 0\\
         hone & 274 & 7 & 0 & 0 & 0 & 0 & 0 & 0 & 0 & 0\\
         Hybrid\_Images & 144 & 19 & 0 & 0 & \cellcolor{c4!10} 9 & \cellcolor{c4!10} \textbf{17} & 0 & \cellcolor{c4!10} 7 & \cellcolor{c4!10} 9 & \cellcolor{c4!10} \textbf{17}\\
         lice & 376 & 25 & 0 & 0 & 0 & \cellcolor{c4!10} \textbf{3} & 0 & 0 & 0 & 0\\
         pso & 168 & 5 & 0 & \cellcolor{c4!10} 1 & \cellcolor{c4!10} \textbf{5} & \cellcolor{c4!10} 1 & 0 & \cellcolor{c4!10} 1 & \cellcolor{c4!10} \textbf{4} & \cellcolor{c4!10} 1\\
         readtime & 284 & 8 & 0 & 0 & \cellcolor{c4!10} 1 & \cellcolor{c4!10} \textbf{2} & 0 & 0 & 0 & \cellcolor{c4!10} \textbf{2}\\
         stocktrends & 384 & 7 & 0 & 0 & \cellcolor{c4!10} 1 & \cellcolor{c4!10} \textbf{3} & 0 & 0 & 0 & 0\\
         TextCNN & 403 & 10 & 0 & 0 & \cellcolor{c4!10} \textbf{7} & \cellcolor{c4!10} 1 & 0 & 0 & \cellcolor{c4!10} 5 & \cellcolor{c4!10} 1\\
         \midrule
         \rowcolor{gray!20}
         \textbf{SUM} & - & 124 & 0 & 25 & 33 & \textbf{52} & 0 & 18 & 41 & \textbf{47} \\
         \rowcolor{gray!20}
         \textbf{AVG$_{weighted}$} & - & - & 0 & 1.72 & 2.85 & \textbf{3.78} & 0 & 1.10 & 2.98 & \textbf{3.05} \\
         \bottomrule
    \end{tabular}
    }
    \label{tab:results_devbench}
    \vspace{-0.3cm}
\end{table}

\begin{table}[t]
    \centering
    \setlength{\abovecaptionskip}{0.1cm}
    \caption{Comparison of \textbf{SketchBLEU} metric between \ourmethod{} and baseline approaches on DevBench.}
    \resizebox{\textwidth}{!}{
    \begin{tabular}{cc|cccc|cccc}
         \toprule
          \multirow{2}{*}{\textbf{Task}} & \multirow{2}{*}{\textbf{\#LOC}}  & \multicolumn{4}{c|}{\textbf{DeepSeek-V3}} & \multicolumn{4}{c}{\textbf{GPT-4o}} \\
          & & \textbf{ChatDev}& \textbf{CodeS} & \textbf{MetaGPT} & \textbf{\ourmethod{}} & \textbf{ChatDev}& \textbf{CodeS} & \textbf{MetaGPT} & \textbf{\ourmethod{}} \\
         \midrule
         ArXiv\_digest & 198 & 30.34 & 52.83 & 94.50 & 94.99 & 49.91 & 50.75 & 95.78 & 96.06\\
         chakin & 162 & 48.81 & 60.01 & 78.16 & 77.61 & 55.17 & 60.43 & 86.33 & 74.55\\
         geotext & 470 & 40.72 & 68.27 & 81.85 & 86.09 & 54.75 & 69.77 & 84.98 & 88.22\\
         hone & 274 & 50.44 & 58.74 & 80.55 & 79.57 & 56.68 & 69.36 & 86.96 & 84.30\\
         Hybrid\_Images & 144 & 30.23 & 48.63 & 95.95 & 94.38 & 56.99 & 40.72 & 95.91 & 95.28\\
         lice & 376 & 38.16 & 70.10 & 92.26 & 91.79 & 56.25 & 78.01 & 93.28 & 93.00\\
         pso & 168 & 41.61 & 67.33 & 92.69 & 89.98 & 54.60 & 54.39 & 92.72 & 90.80\\
         readtime & 284 & 45.41 & 53.94 & 87.60 & 86.20 & 59.24 & 68.02 & 88.70 & 85.50\\
         stocktrends & 384 & 34.36 & 56.64 & 81.07 & 82.52 & 44.22 & 59.84 & 81.85 & 83.05\\
         TextCNN & 403 & 38.26 & 76.14 & 93.43 & 93.46 & 57.43 & 54.06& 94.26 & 93.29\\
         \midrule
         \rowcolor{gray!20}
         \textbf{AVG} & - & 39.83 & 61.26 & 87.81 & 87.66 & 54.52 & 60.54 & 90.01 & 88.41 \\
         \bottomrule
    \end{tabular}
    }
    \label{tab:results_devbench_bleu}
    \vspace{-0.3cm}
\end{table}

Table \ref{tab:results_devbench} presents a comprehensive comparison of \ourmethod{} with three baseline approaches, evaluated using the number of passed test cases on DevBench. 
Except for ChatDev, the code generated by all other approaches is able to pass at least a subset of the test cases.
Under the DeepSeek-V3 setting, \ourmethod{} successfully passes 52 tests, substantially outperforming CodeS (25) and MetaGPT (33).
A similar trend is observed under GPT-4o, where \ourmethod{} leads with 47 passed tests, compared to MetaGPT (41) and CodeS (18). 
Regarding the weighted average score, \ourmethod{} attains 3.78 with DeepSeek-V3 and 3.05 with GPT-4o, consistently demonstrating superior overall performance compared to all baseline approaches.
These results validate the effectiveness of \ourmethod{} in integrating multi-agent collaboration and semantic architectural modeling to produce more complete and executable project implementations.

We further analyzed the reasons for ChatDev's failure to pass any test cases. The primary issue lies in its rigid internal convention of automatically designating \texttt{main.py} as the project's entry file during code generation. 
In practical software development, however, Python projects do not always include a predefined entry point. Many are instead implemented as libraries that provide reusable functionalities for external applications, making an explicit entry file unnecessary. 
Even when an entry file is present, its name is not necessarily \texttt{main.py}. 
This hard-coded convention constrains its flexibility and undermines its generalization capability across diverse types of software projects.
Considering its proposed dataset SRDD, we infer that ChatDev is likely more suitable for handling relatively simple coding tasks with roughly a hundred lines of code, where the requirements are relatively simple and can be described in only a few sentences.

We also investigate the performance on individual tasks.
Firstly, DeepSeek-V3 exhibits superior performance by successfully passing test cases on a larger set of tasks compared to GPT-4o, reflecting notable differences in model capabilities. Secondly, although a subset of tasks remains intractable for all approaches, the majority of tasks with at least some passing test cases are successfully handled by a minimum of two approaches, highlighting the significant influence of task difficulty on performance.
Besides, we also notice that \ourmethod{} and the baseline approaches perform particularly well on the \texttt{ArXiv\_digest} and \texttt{Hybrid\_Images} tasks. We suppose that this pattern is closely related to the number of test cases, as tasks with more test cases tend to have each case assess only a localized portion of the code, which is generally simpler than test cases that evaluate broader sections of the project.

Moreover, we observe that CodeS performs poorly on most tasks but can pass 24 test cases of the \texttt{ArXiv\_digest} task. A possible explanation is that \texttt{ArXiv\_digest} consists of only a single Python file with a simple structure. For CodeS, which uses natural language as an intermediate representation during project generation, more complex requirements and code structures can lead to misinterpretation and parsing errors, resulting in significantly degraded performance.
Regarding the failures of MetaGPT, they are primarily due to its inability to strictly adhere to constraints specified in the documentation. For instance, although the UML class diagram explicitly designates \texttt{Global\_functions} as a pseudo-class for holding global functions, MetaGPT occasionally implements it as an actual class. We also notice cases where MetaGPT generates files that are entirely missing or produces functions and class attributes that are incomplete.
Consequently, these observations further illustrate that directly generating repository structures from user requirements imposes a substantial semantic and structural gap for LLMs, making it difficult to achieve high-quality code outputs.

In addition, we report the results of SketchBLEU in Table \ref{tab:results_devbench_bleu} to evaluate the performance of the generated project code from the perspective of structural similarity.
As shown, MetaGPT and \ourmethod{} achieve highly similar performance under both DeepSeek-V3 and GPT-4o. In contrast, ChatDev and CodeS exhibit significantly lower scores, indicating a substantial gap in structural alignment with the ground truth.
Notably, both MetaGPT and \ourmethod{} also obtained higher numbers of passed tests, suggesting that their generated projects maintain overall coherent structures. However, the close SketchBLEU results between these two approaches imply that, while both capture high-level structural organization effectively, they may still fall short in finer-grained implementation details. This also reflects a limitation of structure-based metrics such as SketchBLEU, which are effective at distinguishing approaches with large structural discrepancies but less sensitive to subtle variations in code correctness or semantic fidelity.

\begin{custommdframed}
\textbf{Answer to RQ1:} On the small-scale project code generation dataset, the majority of existing approaches are capable of producing executable code with partially correct logical behavior, with \ourmethod{} passes the highest number of test cases and demonstrates the best overall performance.
SketchBLEU results indicate that MetaGPT and \ourmethod{} achieve similarly high structural similarity, though it remains limited in capturing fine-grained differences in functional correctness.
\end{custommdframed}
\vspace{0mm}

\subsection{RQ2: Performance on Larger-Scale Projects}

We evaluate the performance of each approach on \ourdata{} to assess their capability in larger-scale project code generation, which contains projects with substantially more lines of code and test cases than those in DevBench. Given that ChatDev has already demonstrated notable limitations in handling small-scale projects, it is excluded from this evaluation. The results are reported in Table \ref{tab:results_ourdata} and Table \ref{tab:results_ourdata_bleu}.

\begin{table}[t]
    \centering
    \setlength{\abovecaptionskip}{0.1cm}
    \caption{Comparison of \textbf{\#Pass} metric between \ourmethod{} and baseline approaches on \ourdata{}.}
    \resizebox{\textwidth}{!}{
    \begin{tabular}{ccc|ccc|ccc}
        \toprule
        \multirow{2}{*}{\textbf{Task}} & \multirow{2}{*}{\textbf{\#LOC}} & \multirow{2}{*}{\textbf{\# Test}} & \multicolumn{3}{c|}{\textbf{DeepSeek-V3}} & \multicolumn{3}{c}{\textbf{GPT-4o}} \\
         & & & \textbf{CodeS} & \textbf{MetaGPT} & \textbf{\ourmethod{}} & \textbf{CodeS} & \textbf{MetaGPT} & \textbf{\ourmethod{}} \\
        \midrule
        bplustree & 1,509 & 356 & 0 & 0 & 0 & 0 & 0 & 0 \\
        cookiecutter &2,805 & 375 & 0 & 0 & 0 & 0 & 0 & 0\\
        csvs-to-sqlite & 816 & 25 & 0 & 0 & \cellcolor{c4!10} \textbf{2} & 0 & 0 & \cellcolor{c4!10} \textbf{2}\\
        deprecated & 597 & 176 & 0 & \cellcolor{c4!10} 35 & \cellcolor{c4!10} \textbf{68} & 0 & \cellcolor{c4!10} 20 & \cellcolor{c4!10} \textbf{23}\\
        djangorestframework-simplejwt & 2,014 & 191 & 0 & 0 & 0 & 0 & 0 & 0\\
        flask & 9,314 & 482 & 0 & 0 & 0 & 0 & 0 & 0\\
        imapclient & 3,531 & 267 & 0 & 0 & 0 & 0 & 0 & 0\\
        parsel & 1,128 & 250 & 0 & 0 & 0 & 0 & 0 & 0\\
        portalocker & 1,958 & 71 & 0 & 0 & 0 & 0 & 0 & 0\\
        pyjwt & 2,690 & 294 & 0 & 0 & 0 & 0 & 0 & \cellcolor{c4!10} \textbf{285}\\
        python-hl7 & 2,434 & 100 & 0 & 0 & 0 & 0 & 0 & 0\\
        rsa & 2,949 & 100 & 0 & 0 & \cellcolor{c4!10} \textbf{90} & 0 & 0 & 0\\
        simpy & 2,184 & 149 & 0 & 0 & 0 & 0 & 0 & 0\\
        tinydb & 2,170 & 204 & 0 & 0 & 0 & 0 & 0 & 0\\
        trailscraper & 890 & 93 & 0 & 0 & 0 & \cellcolor{c4!10} 5 & \cellcolor{c4!10} \textbf{8} & 0 \\
        voluptuous & 3,100 & 161 & 0 & 0 & 0 & 0 & 0 & 0\\
        xmnlp & 1,504 & 23 & 0 & 0 & 0 & 0 & 0 & 0\\
        zxcvbn & 1,402 & 31 & 0 & 0 & 0 & 0 & 0 & 0\\
        \midrule
        \rowcolor{gray!20}
        \textbf{SUM} & - & 3348 & 0 & 35 & \textbf{160} & 5 & 28 & \textbf{310}\\
        \rowcolor{gray!20}
        \textbf{AVG$_{weighted}$} & - & - & 0 & 0.49 & \textbf{7.16} & 0.10 & 0.44 & \textbf{18.19} \\
         \bottomrule
    \end{tabular}
    }
    \label{tab:results_ourdata}
\end{table}

\begin{table}[t]
    \centering
    \setlength{\abovecaptionskip}{0.1cm}
    \caption{Comparison of \textbf{SketchBLEU} metric between \ourmethod{} and baseline approaches on \ourdata{}.}
    \resizebox{\textwidth}{!}{
    \begin{tabular}{cc|ccc|ccc}
        \toprule
        \multirow{2}{*}{\textbf{Task}} & \multirow{2}{*}{\textbf{\#LOC}} & \multicolumn{3}{c|}{\textbf{DeepSeek-V3}} & \multicolumn{3}{c}{\textbf{GPT-4o}} \\
         & & \textbf{CodeS} & \textbf{MetaGPT} & \textbf{\ourmethod{}} & \textbf{CodeS} & \textbf{MetaGPT} & \textbf{\ourmethod{}} \\
        \midrule
        bplustree & 1,509 & 58.57 & 90.97 & 91.06 & 69.13 & 91.07 & 92.81 \\
        cookiecutter &2,805 & 77.80 & 90.50 & 90.80 & 54.83 & 89.70 & 90.89\\
        csvs-to-sqlite & 816 & 67.95 & 90.19 & 87.75 & 73.94 & 90.95 & 91.68\\
        deprecated & 597 & 48.26 & 91.87 & 91.10 & 48.07 & 91.74 & 91.25\\
        djangorestframework-simplejwt & 2,014 & 69.51 & 94.55 & 94.21 & 53.60 & 94.74 & 94.23\\
        flask & 9,314 & 73.92 & 87.98 & 87.17 & 62.08 & 88.95 & 88.88\\
        imapclient & 3,531 & 73.18 & 94.61 & 95.62 & 85.66 & 94.24 & -\\
        parsel & 1,128 & 75.05 & 95.56 & 95.56 & 58.64 & 95.54 & 95.59\\
        portalocker & 1,958 & 70.16 & 90.92 & 92.44 & 68.06 & 89.74 & 91.75\\
        pyjwt & 2,690 & 76.31 & 94.63 & 95.63 & 67.19 & 95.77 & 96.28\\
        python-hl7 & 2,434 & 78.45 & 91.76 & 94.11 & 72.00 & 91.10 & 91.87\\
        rsa & 2,949 & 66.74 & 91.07 & 92.83 & 80.67 & 81.58 & 94.28\\
        simpy & 2,184 & 83.21 & 94.05 & 96.46 & 53.79 & 95.31 & 95.17\\
        tinydb & 2,170 & 75.70 & 93.35 & 95.55 & 52.99 & 95.11 & 95.46\\
        trailscraper & 890 & 70.31 & 94.35 & 92.41 & 79.71 & 92.40 & 92.46\\
        voluptuous & 3,100 & 69.88 & 91.35 & 95.04 & 77.07 & 93.85 & 92.67\\
        xmnlp & 1,504 & 61.92 & 83.20 & 81.63 & 76.16 & 83.80 & 80.95\\
        zxcvbn & 1,402 & 59.75 & 79.22 & 89.27 & 75.35 & 92.18 & 89.67\\
        \midrule
        \rowcolor{gray!20}
        \textbf{AVG} & - & 69.82 & 91.12 & 92.20 & 67.16 & 91.54 & 92.11\\
         \bottomrule
    \end{tabular}
    }
    \label{tab:results_ourdata_bleu}
\end{table}

As shown in Table \ref{tab:results_ourdata}, overall performance decreases notably across all approaches when moving from smalle-scale repositories to more complex and larger projects, reflecting the increased difficulty of maintaining functional correctness at scale. 
Despite this challenge, \ourmethod{} demonstrates clear superiority, achieving 160 passes under DeepSeek-V3 and 310 passes under GPT-4o, substantially outperforming baseline approaches. 

When examining the results at the task level, we observe that among the 18 tasks, only 5 contain any test cases successfully passed by at least one approach, while the remaining 13 tasks show no successful cases across all approaches.
Among the 5 tasks with at least one successful test case, 3 correspond to the only projects in \ourdata{} whose code size is below 1,000 lines.
In addition, \ourmethod{} is the only approach that successfully passes test cases on the \texttt{csvs-to-sqlite} task (816 lines of code), \texttt{rsa} task (2,949 lines of code), and \texttt{pyjwt} project (2,690 lines of code).
This indicates that current project-level code generation approaches still face significant challenges when handling larger-scale projects, and their capability to generate projects exceeding 1,000 lines of code remains limited.

Specifically, the poor performance of CodeS may stem from the same issue observed in RQ1. When handling multi-file projects, using natural language as an intermediate architectural representation tend to suffer from parsing errors, making it difficult to reconstruct the intended architecture and often leading to compilation failures.
Meanwhile, for larger-scale projects, MetaGPT occasionally produce files containing internal agent communication fragments instead of executable code. This suggests that extensive contextual descriptions in larger-scale projects impose substantial comprehension challenges, undermining its ability to maintain consistency and correctness.

Additionally, the results presented in Table \ref{tab:results_ourdata_bleu} for SketchBLEU are consistent with the performance observed on smalle-scale projects.
Even though CodeS fails to pass any test cases, it still achieves average SketchBLEU scores of 69.82 and 67.16, respectively.
Both MetaGPT and \ourmethod{} achieve SketchBLEU scores exceeding 90, even surpassing their performance in RQ1.
On one hand, this suggests that evaluating project code generation performance based solely on structural similarity has inherent limitations, as it may not reliably reflect the actual functional correctness. On the other hand, it also indicates that although the currently generated project code may fail test cases for certain tasks, relatively minor modifications could potentially lead to significantly improved performance. However, in most cases, such fine-grained adjustments remain highly challenging.

\begin{custommdframed}
\textbf{Answer to RQ2:} On \ourdata{}, which contains larger-scale projects, overall performance decreases across all approaches compared to smalle-scale datasets. Despite this, \ourmethod{} achieves the highest number of passed test cases, and is the only approach to succeed on several large projects, highlighting its robustness.
SketchBLEU results also indicate that the generated code may require fine-grained modifications to achieve better performance.
\end{custommdframed}

\subsection{RQ3: Impact of Proposed Components}

\begin{table}[t]
    \centering
    \setlength{\abovecaptionskip}{0.1cm}
    \caption{Impact of key components of \ourmethod{}, reported by the \textbf{\#Pass} metric. \textcolor{green!50!black}{Green} and \textcolor{red!70}{red} numbers beside \#Pass indicate performance changes compared with the approach on the left. Tasks where all approaches failed to pass any test cases are omitted. }
    \resizebox{\textwidth}{!}{
    \begin{tabular}{cc|ccc|ccc}
         \toprule
          \multirow{2}{*}{\textbf{Dataset}} & \multirow{2}{*}{\textbf{Task}} & \multicolumn{3}{c|}{\textbf{DeepSeek-V3}} & \multicolumn{3}{c}{\textbf{GPT-4o}} \\
          & & \textbf{CodeS} & \textbf{\ourmethod{}$_{w/o\ iter}$} & \textbf{\ourmethod{}} & \textbf{CodeS} & \textbf{\ourmethod{}$_{w/o\ iter}$} & \textbf{\ourmethod{}} \\
         \midrule
         \multirow{8}{*}{\textbf{DevBench}} & ArXiv\_digest & 24 & 22 \color{red!70}\small (\downarrowsymbol2) & 25 \color{green!50!black}\small (\uparrowsymbol3) & 10 & 23 \color{green!50!black}\small (\uparrowsymbol13) & 26 \color{green!50!black}\small (\uparrowsymbol3)\\
          & Hybrid\_Images & 0 & 9 \color{green!50!black}\small (\uparrowsymbol9) & 17 \color{green!50!black}\small (\uparrowsymbol8) & 7 & 8 \color{green!50!black}\small (\uparrowsymbol1) & 17 \color{green!50!black}\small (\uparrowsymbol9)\\
          & lice & 0 & 0 & 3 \color{green!50!black}\small (\uparrowsymbol3) & 0  & 0 & 0\\
          & pso & 1 & 0 \color{red!70}\small (\downarrowsymbol1) & 1 \color{green!50!black}\small (\uparrowsymbol1) & 1 & 1 & 1 \\
          & readtime & 0 & 2 \color{green!50!black}\small (\uparrowsymbol2) & 2 & 0  & 0 & 2  \color{green!50!black}\small (\uparrowsymbol2)\\
          & stocktrends & 0 & 1 \color{green!50!black}\small (\uparrowsymbol1) & 3 \color{green!50!black}\small (\uparrowsymbol2) & 0 & 0 & 0\\
          & TextCNN & 0 & 0 & 1 \color{green!50!black}\small (\uparrowsymbol1) & 0 & 0 & 1 \color{green!50!black}\small (\uparrowsymbol1)\\
         \cmidrule{2-8}
          & \cellcolor{gray!20} \textbf{SUM} & \cellcolor{gray!20} 25 & \cellcolor{gray!20} 34 \color{green!50!black}\small (\uparrowsymbol9) & \cellcolor{gray!20} 52 \color{green!50!black}\small (\uparrowsymbol18) & \cellcolor{gray!20} 18 & \cellcolor{gray!20} 32 \color{green!50!black}\small (\uparrowsymbol14) & \cellcolor{gray!20} 47 \color{green!50!black}\small (\uparrowsymbol15) \\
         \bottomrule
         \multirow{6}{*}{\textbf{\ourdata{}}} & csvs-to-sqlite & 0 & 0 & 2 \color{green!50!black}\small (\uparrowsymbol2) & 0 & 2 \color{green!50!black}\small (\uparrowsymbol2) & 2\\
          & deprecated & 0 & 38 \color{green!50!black}\small (\uparrowsymbol38) & 68 \color{green!50!black}\small (\uparrowsymbol30) & 0 & 23 & 23\\
          & pyjwt & 0 & 0 & 0 & 0 & 0 & 285 \color{green!50!black}\small (\uparrowsymbol285)\\
          & rsa & 0 & 0 & 90 \color{green!50!black}\small (\uparrowsymbol90) & 0 & 0 & 0\\
          & trailscraper & 0 & 0 & 0 & 5 & 0 \color{red!70}\small (\downarrowsymbol5) & 0 \\
         \cmidrule{2-8}
         & \cellcolor{gray!20} \textbf{SUM} & \cellcolor{gray!20} 0 & \cellcolor{gray!20} 38 \color{green!50!black}\small (\uparrowsymbol38) & \cellcolor{gray!20} 160 \color{green!50!black}\small (\uparrowsymbol122) & \cellcolor{gray!20} 5 & \cellcolor{gray!20} 25 \color{green!50!black}\small (\uparrowsymbol20) & \cellcolor{gray!20} 310 \color{green!50!black}\small (\uparrowsymbol285)\\
         \bottomrule
    \end{tabular}
    }
    \label{tab:ablation}
    \vspace{-0.3cm}
\end{table}

In this RQ, we analyze the impacts of two key components of \ourmethod{}: the SSAT and the iterative code optimization mechanism.
To support the comparison, we additionally conducted an experiment in which we retained the same generation agents as in \ourmethod{} but removed all judging agents, denoted as \ourmethod{}$_{w/o\ iter}$, which does not employ iterative refinement guided by judge feedback.
Specifically, \textbf{we compare \ourmethod{}$_{w/o\ iter}$ with CodeS to assess the effect of SSAT}, as CodeS follows a three-stage process of repository sketch, file sketch, and function body. In contrast, \ourmethod{}$_{w/o\ iter}$ replaces the repository sketch with the generation of SSAT, which then guides subsequent skeleton and code generation.
Besides, \textbf{we compare \ourmethod{}$_{w/o\ iter}$ with \ourmethod{} to assess the effect of the iterative code optimization mechanism.}
To ensure meaningful evaluation, we select tasks for which at least one of the evaluated approaches can pass the test cases in this experiment.

The results are shown in Table \ref{tab:ablation}, with 12 out of 28 tasks successfully passed the tests by at least one approach.
Compared with CodeS, \ourmethod{} achieves a overall increase in the total number of passed tests, especially in larger-scale dataset \ourdata{}, demonstrating the effectiveness of the proposed SSAT. Moreover, \ourmethod{} yields significantly more successful test cases than \ourmethod{}$_{w/o\ iter}$, further confirming the importance of the iterative code optimization mechanism in improving generation quality.

We also noticed that in some tasks, \ourmethod{}$_{w/o\ iter}$ passes fewer test cases compared to CodeS.
To identify the potential reason, we further analysis the corresponding source code.
In \texttt{trailscraper} task, \ourmethod{}$_{w/o\ iter}$ encapsulated functions across files but failed to implement the imported ones, as it strictly followed the user requirements. This reveals that ensuring inter-file consistency is a key challenge in larger-scale project generation, which requires stricter evaluation.
In \texttt{pso} task, the PRD did not specify whether initialization should use fixed values. \ourmethod{}$_{w/o\ iter}$ adopted random initialization, while CodeS used fixed values. 
In \texttt{ArXiv\_digest} task, \ourmethod{}$_{w/o\ iter}$ assigned an incorrect return type to a function and omitted certain boundary condition handling.
These observations indicate that iterative optimization is crucial for handling inter-file dependencies, ambiguous requirements, and functional correctness, enhancing the reliability of generated project-level code.
However, in terms of quality and style, \ourmethod{}$_{w/o\ iter}$ and \ourmethod{} consistently produce more encapsulated code than CodeS, enhancing conciseness and readability.

\begin{custommdframed}
\textbf{Answer to RQ3:} 
The two key components of \ourmethod{} both make important contributions to the overall performance. Specifically, SSAT enables the LLM to better understand user requirements, and iterative optimization helps facilitates the identification and correction of errors across files, ensuring better inter-file consistency and functional completeness. Together, these components significantly enhance the reliability, readability, and overall quality of the generated project-level code.
\end{custommdframed}

\section{Discussion}

\subsection{Quantitative Analysis and Error Diagnosis}

To gain deeper insights into the quantitative analysis and error type for failure of project code generated by \ourmethod{}, we conduct a more detailed investigation using the code generated by DeepSeek-V3.

\begin{figure}[t]
  \centering
  \begin{subfigure}[t]{0.9\linewidth}
    \centering
    \includegraphics[width=\linewidth]{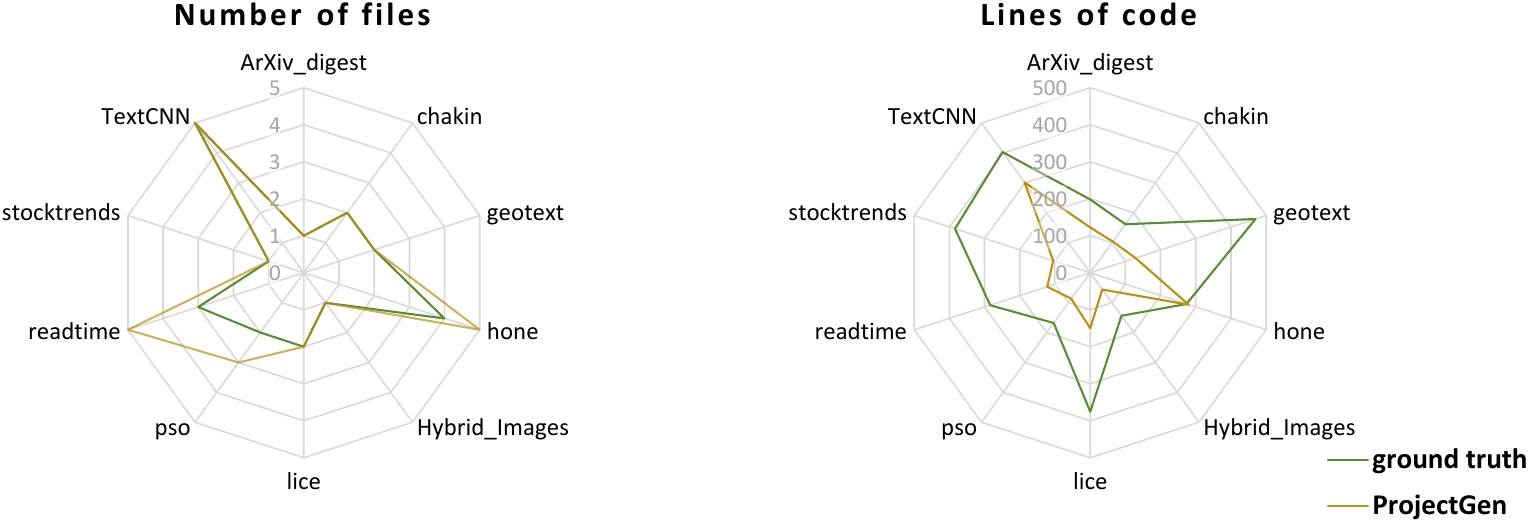}
    \caption{Comparison on small-scale project-level code generation dataset DevBench.}
    \label{fig:devbench_a}
  \end{subfigure}
  \hfill
  \begin{subfigure}[t]{0.9\linewidth}
    \centering
    \includegraphics[width=\linewidth]{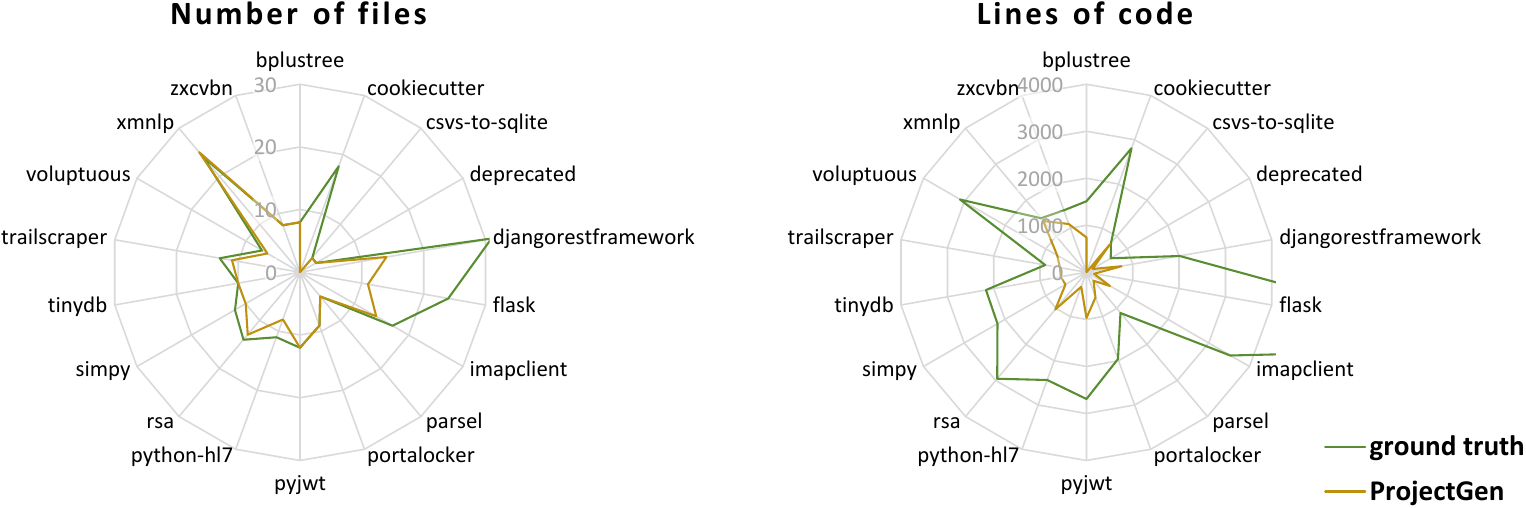}
    \caption{Comparison on larger-scale project-level code generation dataset \ourdata{}.}
    \label{fig:devbench_b}
  \end{subfigure}
  \caption{Comparison of project size between ground truth and code generated by \ourmethod{} using DeepSeek-V3.}
  \label{fig:analysis}
  \vspace{-0.3cm}
\end{figure}

\subsubsection{Project Scale.}
We first evaluate the number of files and total lines of code in the generated projects, which reflect the project scale and structural characteristics, indicating how closely it resembles real-world implementations.
The results are provided in Figure \ref{fig:analysis}.

For the small-scale code generation dataset DevBench, the projects generated by \ourmethod{} show no significant difference in the number of files compared with the ground truth, and in some cases even slightly exceed it. This can be explained by the observation that certain \texttt{\_\_init\_\_.py} in the ground truth are empty, whereas \ourmethod{} produces meaningful content within them, leading to an increase in the number of files containing executable code.
In terms of lines of code, although \ourmethod{} generally produces fewer lines than the ground truth, the differences are not substantial. It is also notable that on the \texttt{Hybrid\_Images} task, \ourmethod{} generates code with only one-third of the ground truth’s scale, yet successfully passes 17 out of 19 test cases.
We also calculated the cyclomatic complexity of the generated code, and the average value for DevBench code generated by \ourmethod{} is 2.44.
Compared to the ground truth value reported in Table \ref{tab:benchmark_comparison}, the discrepancy may stem from the variations in code implementation.

For larger-scale datasets, the gap in project size between \ourmethod{} and the ground truth becomes increasingly pronounced.
For projects containing more than 15 files, \ourmethod{} struggles to reproduce a comparable number of files. In terms of lines of code, the generated projects predominantly remain around the scale of one thousand lines.
The cyclomatic complexity of the generated code is 2.72, which does not exhibit a notable difference with ground truth.
Although code scale does not necessarily reflect code quality or functional correctness, it nevertheless suggests that generating complete project-level code with LLMs still faces notable challenges, such as effectively guiding the LLM to produce code of larger or more appropriate scale.

\begin{figure}[t]
  \centering
  \includegraphics[width=0.8\linewidth]{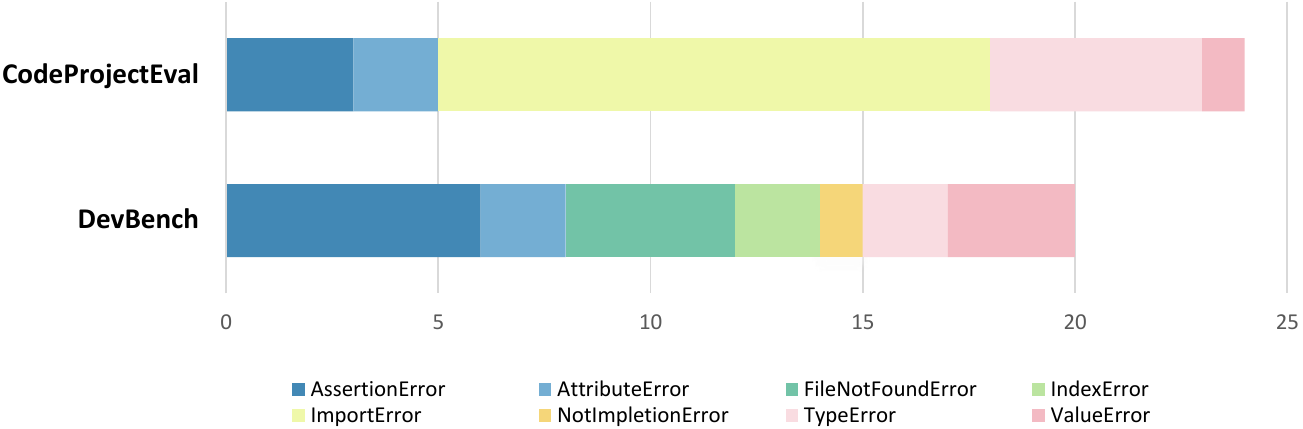}
  \caption{Comparison of error types in code generated by \ourmethod{} on DevBench and \ourdata{}.}
  \label{fig:error_analysis}
  \vspace{-0.3cm}
\end{figure}

\subsubsection{Error Type.}
We further analyze the potential reasons for errors. For code generated by \ourmethod{}, we collect the types of errors reported after running unit tests. If a particular error type occurs multiple times within the same task, it is recorded only once. The results can be seen in Figure \ref{fig:error_analysis}.

In DevBench, errors primarily arise from implementation-level details. \texttt{AssertionError} indicates that the execution results do not match the ground truth, while \texttt{AttributeError}, \texttt{TypeError}, \texttt{IndexError}, and \texttt{ValueError} generally reflect incorrect handling or configuration of specific variables. Additionally, \texttt{FileNotFoundError} occurs when the code attempts to access a file without a correctly specified path, and \texttt{NotImplementedError} arises when certain parts of the code skeleton remain unfilled during the code filling process.

However, in \ourdata{}, the majority of errors stem from \texttt{ImportError}, indicating that issues arise at the very initial import stage, often before the internal logic of the code is even evaluated. In contrast, DevBench does not exhibit \texttt{ImportError}. This discrepancy can be attributed to two factors: firstly, projects in \ourdata{} are more complex, and when handling intricate requirements and code, the LLM is more likely to produce inconsistencies such as undefined functions or mismatched names; secondly, during iterative code refinement, each round of error feedback typically reports only the first encountered error. As a result, even after completing all iterations, the internal logic of the code may remain uncorrected. Extending the number of iterations could to some extent address this issue but would incur substantial overhead, which highlights a key challenge in generating code for larger-scale projects.

\subsection{Case Study}

\begin{figure}[t]
  \centering
  \includegraphics[width=0.85\linewidth]{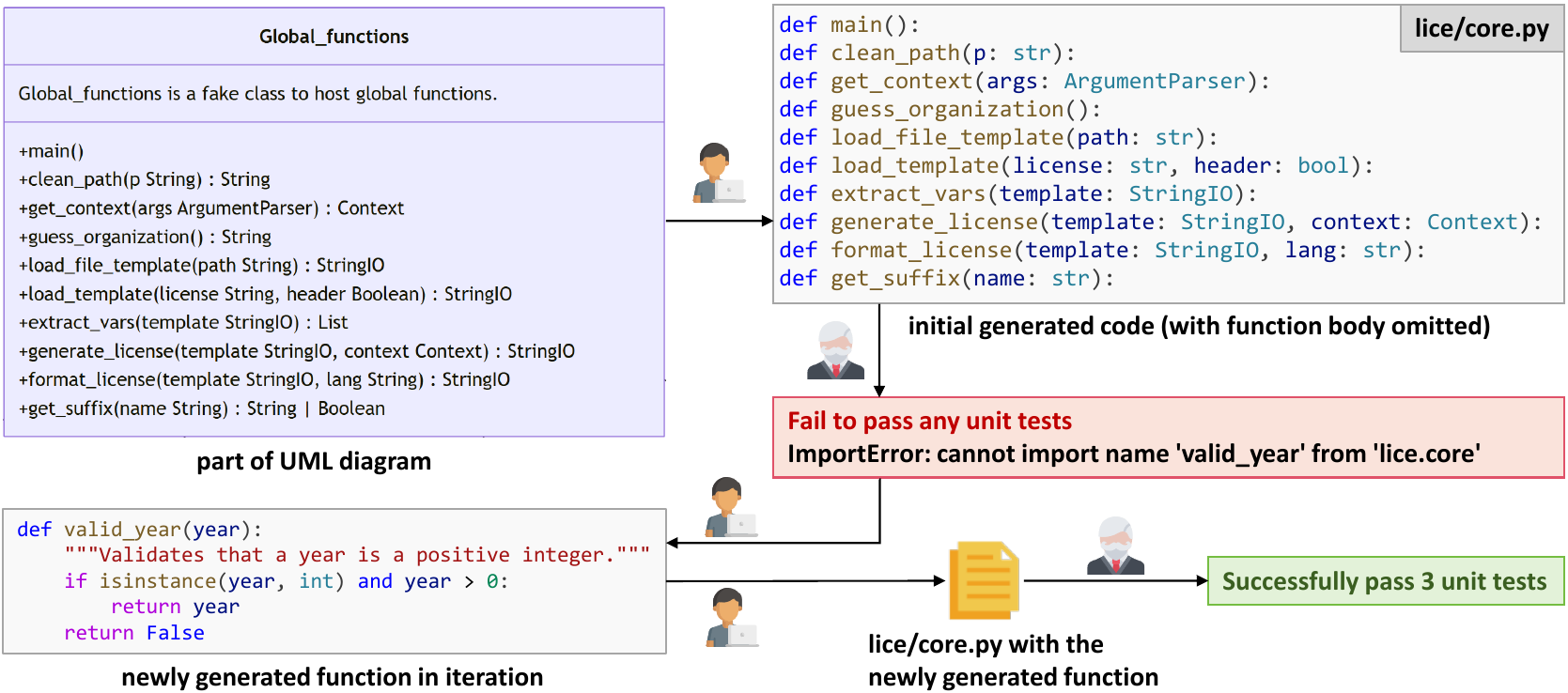}
  \caption{Case Study of the \texttt{lice} task in DevBench.}
  \label{fig:case}
  \vspace{-0.3cm}
\end{figure}

To provide a more detailed illustration of \ourmethod{}’s performance, we present a case study in Figure \ref{fig:case}.
In this task, \ourmethod{} first generated the corresponding global functions strictly according to the provided UML diagrams. However, the evaluation feedback from \textsc{JudgeC} indicated that the \texttt{valid\_year} function was missing.
Based on this feedback, \textsc{CodeAgent} regenerated the implementation of the missing function and incorporated it into the corresponding code file. Subsequent testing showed that the updated implementation successfully passed three unit tests.
A careful examination revealed that this function was not included in the user requirements, which is similar to the common situations encoutered in real-world software development where errors or omissions in the design stage are discovered only during code implementation.
This demonstrates that execution feedback based on tests can help identify aspects that were not covered in the initial user requirements during the implementation phase.

Furthermore, when examining the code generated by the baseline approaches on this task, we observed that CodeS did not strictly adhere to the requirements, with some required functions missing and several undefined functions present. 
Similarly, MetaGPT also failed to define some global variables that were required.
This observation highlights the inherent difficulty of directly translating extensive requirement specifications into code structures using LLMs. By contrast, \ourmethod{}, leveraging the SSAT, effectively mitigates this challenge by providing an intermediate architectural representation that guides the code generation process and ensures closer alignment with the intended system design.

\subsection{Threats to Validity}

The \textbf{internal threat to validity} mainly lies in the construction of \ourdata{}, as some of the included repositories may have been studied by LLMs during pretraining.
However, results reported in \cite{zhaocommit0} indicate that, even when LLMs are provided with complete code skeletons and all test cases used for evaluation, the generated code still performs poorly, passing only a small fraction of the tests. Generating a complete project directly from user requirements is even more challenging. Therefore, the potential data leakage in \ourdata{} may not have a significant impact on the results.
Besides, we incorporate an iterative refinement process guided by check test execution feedback during project-level code generation. However, the check tests we construct are completely disjoint from the unit tests used for final evaluation, ensuring no data leakage.
In real-world software engineering, before a project undergoes systematic testing, developers also rely on small-scale test cases to identify and resolve issues in the code.

The \textbf{external threat to validity} mainly stems from dataset limitations, as we perform evaluation on a limited set of datasets, and all experiments are conducted on Python projects.
Thus, the results obtained may not generalize to other experimental settings.
However, there is a limited availability of project-level code generation datasets that provide executable test cases for automated evaluation, restricting opportunities for evaluation across diverse projects.
Furthermore, our newly proposed dataset \ourdata{} exhibits variation in both the number of files and lines of code per project, enabling more comprehensive and diverse evaluation.
In addition, \ourmethod{} is inherently language-agnostic, and its underlying principles can be applied to other programming languages.

\section{Conclusion}

In this paper, we focus on project-level code generation task, which requires the implementation of project code directly from user requirements.
We first construct \ourdata{}, a dataset constructed from 18 real-world repositories, with larger-scale projects compared to existing datasets, supplemented with documentation and executable test cases to support rigorous automated evaluation.
We further propose \ourmethod{}, a multi-agent framework that decomposes the project generation process into architecture design, skeleton generation, and code filling stages, enhanced with iteration-based refinement and memory-augmented context management to handle hierarchical dependencies and maintain code quality.
Experimental results demonstrate that \ourmethod{} achieves state-of-the-art performance, significantly outperforming existing approaches on both small-scale and larger-scale tasks.

In addition, we consider several future directions that may further boost the performance of \ourmethod{}. 
Given the poor performance of current project-level code generation approaches on larger-scale projects, we suppose that incorporating human interaction is essential for guiding and improving the generation process.
In this context, we aim to explore interactive modification of software architecture representations, enabling human developers to directly guide and adjust generated architectures in real time, thereby increasing both the fidelity and controllability of the generated projects.
Besides, since SSAT provides a structured representation and early-stage architectural designs are often modified or refactored during project development, we envision that generated code could be leveraged to iteratively update and refine the architectural representation, thereby establishing a tighter feedback loop between implementation and design, which may improve the overall coherence and accuracy of generated projects.
Moreover, we also plan to extend \ourdata{} to support the evaluation and development of project-level code generation on more programming languages.



\bibliographystyle{ACM-Reference-Format}
\bibliography{references}



\end{document}